\documentclass[a4paper,11pt]{article}
\pdfoutput=1
\usepackage{jcappub}
\usepackage[T1]{fontenc}
\usepackage[utf8]{inputenc}

\author[1,2]{Chris Clarkson,}
\author[2,3]{Roy~Maartens}

\affiliation[1]{Centre for Theoretical Physics \& Astronomy, Queen Mary University of London, London E1 4NS, United Kingdom}
\affiliation[2]{Department of Physics \& Astronomy, University of the Western Cape, Cape Town 7535, South Africa}
\affiliation[3]{National Institute for Theoretical \& Computational Sciences, Cape Town 7535, South Africa}

\emailAdd{chris.clarkson@qmul.ac.uk}
\emailAdd{rmaartens@uwc.ac.za}

\usepackage{graphicx}
\usepackage{subcaption}
\usepackage{needspace}
\usepackage{flafter}
\usepackage{amssymb}
\usepackage{amsmath}
\usepackage{bm}
\usepackage{mathtools}
\usepackage{enumerate}
\usepackage{natbib,hyperref}
\usepackage{xcolor}

\newcommand{\ud}{\mathrm{d}}
\newcommand{\cd}{\mathrm{D}}

\def\si{\sigma}
\def\om{\omega}

\def\w{{\rm w}}

\title{Beyond dust in local cosmography: perturbative Vlasov predictions}

\abstract{The dust approximation used in cosmography neglects the velocity dispersion
generated during structure formation. This dispersion produces isotropic and
anisotropic pressure, whose gradients accelerate the mean matter flow even
though individual particles follow geodesics. The acceleration introduces an
intrinsic dipole into the local distance--redshift relation. We calculate this
dipole and the accompanying changes in expansion and shear using linear Vlasov
perturbation theory about an FLRW background, with a mean dispersion history
motivated by halo estimates. The assumed FLRW background prevents an independent test of the
cosmological principle; the numerical predictions also depend on the adopted
$\Lambda$CDM cosmology. Pressure also changes the growth of structure,
modifying the shear and the spatial gradients that govern how the distance
anisotropy varies with redshift. The largest correction at the lowest order in redshift is to the shear,
while changes in the expansion gradient can produce a larger dipole at the
next order. The resulting intrinsic dipole is far smaller than the observed
low-redshift dipole, whose behaviour nearest the observer is dominated by
motion between the matter and CMB frames.}

\begin{document}
\maketitle
\pagenumbering{roman}
\setcounter{tocdepth}{2}
\tableofcontents
\clearpage
\pagenumbering{arabic}
\allowdisplaybreaks
\setlength{\emergencystretch}{1em}
\widowpenalty=10000
\clubpenalty=10000
\displaywidowpenalty=10000
\renewcommand{\topfraction}{0.9}
\renewcommand{\textfraction}{0.08}
\setcounter{topnumber}{3}

\section{Introduction}
\label{sec:introduction}

The dust model assigns a unique matter velocity to each event and
describes the flow well while neighbouring streams remain separate. As
structure forms, particle streams falling into the same region pass through
one another, and different velocities coexist at the same position.
The resulting velocity dispersion produces isotropic and anisotropic
pressure through the random particle motions, even though the
particles do not collide. The generation and evolution of these
pressures have been studied using velocity moments and the phase-space
distribution itself
\cite{Buchert:1997dr,Valageas:2001df,Pueblas:2008uv,
Erschfeld:2018zqg,Erschfeld:2020blf,Erschfeld:2023aqr,
Nascimento:2024hle,Heisler:2025zlc}.
Individual dark-matter particles follow geodesics, but gradients of
these pressures can accelerate their mean flow. Our companion paper,
hereafter Paper I \cite{ClarksonMaartens:CovariantCompanion}, derives
this acceleration without assuming a background spacetime and shows
how it enters local cosmography. Here we estimate its magnitude for a prescribed mean dispersion,
together with the associated changes in expansion and shear.

The dust-based covariant cosmographic analysis of
\cite{Kalbouneh:2025jnp} uses sources in the range $0.01<z<0.1$ from
the Cosmicflows-4 catalogue \cite{Tully:2023Cosmicflows4}. Extending
the fit to include sources below $z\simeq0.01$ produces systematic
shifts in the fitted parameters and worsens the goodness of fit.
Reference~\cite{Kalbouneh:2025jnp} associates this sensitivity with the limits
of representing nearby galaxy motions by a smooth fluid.
This motivates examining the pressure terms omitted by
the dust model, beginning with their effects on the local cosmographic
coefficients. We calculate the local coefficients here; we do not make the full
distance prediction needed to refit the catalogue.

Covariant cosmography expresses the distance--redshift relation in
terms of the local expansion, shear and 4-acceleration
\cite{Kristian:1966,MacCallum:1970,Clarkson:1999zq,Heinesen:2020bej,
Maartens:2023tib,Pasten:2026rgb}. Vlasov perturbation theory (VPT)
\cite{Garny:2022tlk,Garny:2022kbk,Garny:2025ovs,Garny:2025zlq}
describes the density, mean velocity and velocity cumulants in a perturbed
FLRW universe, from which the kinematic quantities follow. The mean
dispersion is generated by nonlinear structure formation; VPT evolves its
spatial fluctuations. We use the halo-based dispersion history of
\cite{Garny:2025zlq} as an input, with its relation to the required spatial
mean discussed in section~\ref{sec:vpt-definition}.

We evolve linear fluctuations about this mean, compare successive cumulant
truncations and identify the scales contributing after spatial smoothing.
At zeroth order in redshift, the largest correction is a $0.3$--$1.3\%$
change in shear relative to dust on the scales considered here. The
acceleration dipole is much smaller. Changes in the expansion and shear
gradients enter at the next redshift order, where the growth-induced dipole
can exceed the shear correction. We calculate these contributions and their
variation between observers, then relate them to the coefficients used in
dust cosmographic fits.

\section{Collisionless matter and local cosmography}
\label{sec:covariant-start}

We follow the terminology and covariant notation of Paper I
\cite{ClarksonMaartens:CovariantCompanion}. The matter energy frame is defined by a vanishing energy flux,
$q_a=0$. Let $u^a$ be its 4-velocity, $h_{ab}=g_{ab}+u_au_b$ the spatial
projector and $\cd_a$ the projected spatial derivative. An overdot
denotes differentiation along $u^a$, with free indices projected when
indicated by $\perp$. Angle brackets on tensor indices denote the
projected, symmetric and tracefree (PSTF) part. To leading non-relativistic order, the isotropic pressure $P$ and
anisotropic pressure $\pi_{ab}$ are density-weighted second moments
of particle velocities about the local mean. We set $c=1$ in covariant equations and restore it in
numerical expressions.

In this frame, the exact momentum conservation law of Paper I gives
\begin{align}
 (\rho+P)A_a+\cd_aP+\cd^b\pi_{ab}+A^b\pi_{ab}=0,
 \label{eq:vpt-cov-A}
\end{align}
where $\rho$ is the total matter energy density and
$A_a=u^b\nabla_bu_a$. If $m$ is the particle mass and $n$ the
number density, the rest-mass density through second order in
particle speed is $\rho_m=mn=\rho-3P/2+O(v^4\rho)$, so that
\begin{align}
 A_a\simeq-\rho_m^{-1}\big(\cd_aP+\cd^b\pi_{ab}\big).
 \label{eq:main-A-covariant}
\end{align}
The acceleration of the mean flow is therefore fixed by gradients of
its isotropic and anisotropic pressure. The energy and particle-number frames agree at this
order, since their relative velocity first appears at third velocity
order \cite{ClarksonMaartens:CovariantCompanion}. Thus either frame gives the same acceleration to the order used here.

At second velocity order, Paper I gives the energy and pressure equations
\begin{align}
 \dot\rho+(\rho+P)\Theta+\si^{ab}\pi_{ab}&=0,
 \label{eq:vpt-cov-rho}\\
 \dot P+\frac53\Theta P+\frac23\si^{ab}\pi_{ab}&=0,
 \label{eq:vpt-cov-P}\\
 \big(\dot\pi_{ab}\big)_\perp+\frac53\Theta\pi_{ab}
 +2\pi^c{}_{\langle a}\big(\si_{b\rangle c}+\om_{b\rangle c}\big)
 +2P\si_{ab}&=0.
 \label{eq:vpt-cov-pi}
\end{align}
Here $\Theta$, $\si_{ab}$ and $\om_{ab}$ are the volume expansion,
shear and vorticity. Expansion dilutes the pressure and shear generates
anisotropic pressure from an initially isotropic dispersion. These
equations contain no mechanism for generating dispersion from exactly
cold single-stream initial data. A statistical description after stream
crossing must supply the dispersion produced by the unresolved dynamics;
we specify this input in the next section. The relation to the moment
hierarchy, including the third moments omitted here, is derived in
section~\ref{sec:second-moment-evolution}.

The directional Hubble parameter is
\cite{Kristian:1966,Maartens:2023tib,ClarksonMaartens:CovariantCompanion}
\begin{align}
 \mathbb H(n^a)=\frac13\Theta-A_an^a+\si_{ab}n^an^b,
 \label{hub}
\end{align}
where $n^a$ points from the observer towards the source, with
$n^an_a=1$ and $u_an^a=0$, as in Paper I. The
intrinsic Hubble dipole is $\mathbb H_a=-A_a$ and the quadrupole is
$\mathbb H_{ab}=\si_{ab}$. To see how these enter a distance measurement,
write the local luminosity-distance expansion
\cite{Kristian:1966,Heinesen:2020bej,ClarksonMaartens:CovariantCompanion} as
\begin{align}
 d_L(z,n^a)=\frac{z}{\mathbb H(n^a)}
 \left[1+\frac12\big(1-\mathbb Q(n^a)\big)z+O(z^2)\right],
 \label{eq:main-distance-expansion}
\end{align}
which defines the directional deceleration parameter $\mathbb Q$.
Following \cite{Kalbouneh:2025jnp}, we describe the anisotropy of $d_L$ by
\begin{align}
 \eta(z,n^a)=\log_{10}\frac{z}{d_L(z,n^a)}
 -\left\langle\log_{10}\frac{z}{d_L(z,n^a)}\right\rangle_{\bm n},
 \label{eq:main-eta-def}
\end{align}
where $\langle\cdots\rangle_{\bm n}$ is the sky average. Expanding the
logarithm to first order about FLRW gives
\begin{align}
 (\ln10)\eta(z,n^a)
 =\left[-\frac{A_a}{H}n^a+\frac{\si_{ab}}{H}n^an^b\right]_{\rm o}
 +O(z,\delta_m^2).
 \label{eq:main-eta-A}
\end{align}
Both terms are evaluated at the observer and are independent of source
redshift at leading order. The acceleration contribution is an intrinsic
property of the matter congruence. Motion of the observer relative to
that frame adds a Doppler dipole whose leading low-redshift amplitude
scales as $v_\parallel/(cz)$, where $v_\parallel$ is the observer's
velocity relative to the matter frame projected along $n^a$
\cite{Maartens:2023tib,ClarksonMaartens:CovariantCompanion}.
A constant intrinsic term cannot be represented by a single change in
the observer velocity over a range of redshifts, although the two can
be degenerate in a narrow shell. Holding the smoothing radius and observer velocity fixed, the
ratio of the intrinsic dipole to the Doppler dipole decreases in proportion
to $z$.

These predictions refer to the mean matter flow. Individual galaxies
can be approximately geodesic without the mean velocity of a
multi-stream galaxy population defining a geodesic congruence. A
comparison with a galaxy catalogue therefore also requires the relation
between the matter and tracer frames (section~\ref{sec:galaxy-tracers}).

\section{Vlasov perturbation theory}
\label{sec:vlasov-acceleration}
\subsection{Velocity moments and mean dispersion}
\label{sec:vpt-definition}

The Vlasov equation follows the distribution of collisionless
particles in position and momentum. Taking velocity moments converts
it into equations for density, mean velocity, dispersion and higher
moments. These equations form a hierarchy because transport of a
moment depends on the next one: for example, the pressure equation
contains the divergence of the third velocity moment. A finite set of
fluid equations therefore requires a prescription for the omitted
moments.

VPT uses velocity cumulants to describe the same distribution in the
non-relativistic, subhorizon limit
\cite{Garny:2022tlk,Garny:2025ovs}. Let $\bm v_p$ be a particle's
peculiar velocity and $\bm v=\langle\bm v_p\rangle_{\bm x}$ its
mean at fixed position and time. This average integrates over the local
velocity distribution, weighting each stream by its particle density.\footnote{For
any particle quantity $X$, the local average is
$\langle X\rangle_{\bm x}=\int\ud^3p\,{\sf f}_{\rm ps}X/
\int\ud^3p\,{\sf f}_{\rm ps}$, with $\bm x$ and $\tau$ held fixed and
$\bm v_p=\bm p/a$. Here $\bm p$ is comoving momentum per unit mass;
${\sf f}_{\rm ps}$ is defined in appendix~\ref{app:velocity-cumulants}.
There is no spatial smoothing in this average.}
The second cumulant is the dispersion about that local mean,
\begin{align}
 C_{ij}=\left\langle(v_{p,i}-v_i)(v_{p,j}-v_j)\right\rangle_{\bm x},
 \qquad \rho_m C_{ij}=P\delta_{ij}+\pi_{ij}.
 \label{eq:main-second-cumulant}
\end{align}
The cumulant cutoff $c_{\max}$ specifies the highest velocity rank
whose fluctuations are evolved. At $c_{\max}=2$ we retain density,
mean velocity and all six components of $C_{ij}$, setting higher
cumulants to zero. At this order the resulting equations agree with
the non-relativistic limit of Paper I's second-velocity-order closure
\cite{ClarksonMaartens:CovariantCompanion}.
At $c_{\max}=3$ we also evolve the third cumulant, which describes
the asymmetry of the local velocity distribution and transports
dispersion. The relation between cumulants and central moments is given in
appendix~\ref{app:velocity-cumulants}.

We evolve linear fluctuations about homogeneous mean cumulants in an
FLRW spacetime, using the Vlasov--Poisson equations on subhorizon scales.
The perturbative order and the cumulant cutoff are separate choices:
each evolved cumulant is treated at linear order. The nonzero mean
dispersion represents random particle motions generated by nonlinear
collapse. Standard pressureless perturbation theory
\cite{Bernardeau:2002PT} sets all cumulants above the mean velocity to zero.

We use $\mathcal H=aH=a'/a$, where a prime denotes a derivative with
respect to conformal time $\tau$, the pressureless growing mode $D(a)$
with $D(1)=1$, and $f=\ud\ln D/\ud\ln a$. Statistical isotropy makes the
mean dispersion tensor proportional to $\delta_{ij}$, with amplitude
\begin{align}
 \epsilon(a)=\frac{1}{3(\mathcal Hf)^2}
 \left\langle\left\langle|\bm v_p-\bm v|^2\right\rangle_{\bm x}
 \right\rangle_{\rm ens},
 \qquad k_\sigma(a)=\epsilon(a)^{-1/2}.
 \label{eq:main-epsilon}
\end{align}
The inner average is over particle momenta at fixed position, as in
eq.~\eqref{eq:main-second-cumulant}. The outer average is over statistical
realisations of the resulting local dispersion field; for a statistically
homogeneous, ergodic ensemble, it can also be estimated by a spatial
volume average.
The normalised dispersion $\epsilon$ has dimensions of comoving
length squared, with
$C_{ij}=(\mathcal Hf)^2\epsilon\delta_{ij}+\delta C_{ij}$ and
$(\mathcal Hf)^2\epsilon=\langle P/\rho_m\rangle_{\rm ens}$ at
the retained velocity order. The scale $k_\sigma$ marks where
dispersion significantly changes the evolution of perturbations.
When redshift is used as the argument, we write
$\epsilon(z)\equiv\epsilon[a=(1+z)^{-1}]$, and likewise for $k_\sigma$.

In an FLRW spacetime without dispersion production, random peculiar
velocities of non-relativistic particles redshift as $a^{-1}$,
giving $C_{ij}\propto a^{-2}$.
Together with rest-mass conservation, $\rho_m\propto a^{-3}$,
this gives $P\propto a^{-5}$, the FLRW limit of
eq.~\eqref{eq:vpt-cov-P}. In our normalisation the freely
redshifting solution obeys $(\mathcal Hf)^2\epsilon\propto a^{-2}$.
Nonlinear collapse and stream crossing generate further random
motion, so the mean dispersion history must also include its production.

We prescribe $\epsilon(a)$ using the fitted halo estimate of
\cite{Garny:2025zlq}. Halo density and velocity-dispersion profiles,
combined with the halo mass function, supply an estimate of the random
motions generated by nonlinear structure formation. The fiducial
prescription includes halo outskirts out to $1.5r_{\rm vir}$ and gives
$k_\sigma(0)=0.36\,h\,{\rm Mpc}^{-1}$. We use this history as the
background dispersion in the linear VPT equations; its construction is
described in appendix~\ref{app:vpt-epsilon}.

We use the calibration cosmology of \cite{Garny:2025zlq}: flat
$\Lambda$CDM with $\Omega_{m0}=0.25$, $\Omega_{b0}=0.04$, $h=0.7$,
$\sigma_8=0.8$ and $n_s=1$. These are the parameters used to obtain the
dispersion history. They differ from the observational best-fit values
from Planck \cite{Planck:2018Parameters}. A different cosmology changes the
density spectrum, growth and dispersion history together.

\Needspace{11\baselineskip}
\subsection{Second-moment evolution}
\label{sec:second-moment-evolution}

To connect the dispersion tensor to the covariant hierarchy of Paper I,
write its moments as $\Pi^{(n)}_{{\cal A}_\ell}$, where
${\cal A}_\ell=a_1\cdots a_\ell$ labels the angular rank and $n$ counts
additional factors of squared particle speed. The velocity order is
$m=\ell+2n$, and the lowest moments are
\begin{align}
 \Pi=\rho,\qquad \Pi_a=q_a,\qquad
 \Pi^{(1)}=3P,\qquad \Pi_{ab}=\pi_{ab}.
 \label{eq:main-covariant-moment-identities}
\end{align}
Here $\Pi_{{\cal A}_\ell}\equiv\Pi^{(0)}_{{\cal A}_\ell}$.
The second-velocity-order closure keeps $\ell+2n\leq2$; the rank alone
does not specify the velocity order
\cite{Lewis:2002nc,ClarksonMaartens:CovariantCompanion}.
These momentum-space ranks are distinct from the sky multipoles.
Dividing by the rest-mass density gives

\begin{align}
 M^{(n)}_{{\cal A}_\ell}
 \equiv\frac{\Pi^{(n)}_{{\cal A}_\ell}}{\rho_m}
 =\left\langle v^{2n}v_{\langle a_1}\cdots v_{a_\ell\rangle}\right\rangle_{\bm x}+O(v^{m+2}),
 \qquad
 \Pi^{(n)}_{{\cal A}_\ell}=O\big(v^{\ell+2n}\rho\big).
 \label{eq:vpt-normalised-moments}
\end{align}

The unnormalised $\Pi^{(n)}_{{\cal A}_\ell}$ are exactly the moments
of Paper I; $M^{(n)}_{{\cal A}_\ell}$ denotes their ratio to $\rho_m$.
Here $v^a$ is particle velocity measured in the matter frame, so these
are central velocity moments at leading non-relativistic order.
The physical spatial components of the second moment $M_{ab}$ are the
dispersion tensor $C_{ij}$ of eq.~\eqref{eq:main-second-cumulant}.

At leading non-relativistic order in each moment, we divide the
covariant hierarchy by $\rho_m$ and use
rest-mass conservation, $\dot\rho_m+\rho_m\Theta=0$. The unnormalised
second-moment equations \eqref{eq:vpt-cov-P}--\eqref{eq:vpt-cov-pi}
have damping $5\Theta/3$; dividing by $\rho_m$ reduces it to $2\Theta/3$.
In general, $\tfrac13(\ell+3+2n)\Theta$ becomes $\tfrac13(\ell+2n)\Theta$,
and a normalised moment of velocity order $m=\ell+2n$ obeys
\begin{align}
 \dot M+\frac{m}{3}\Theta\,M=\text{sources}
 \qquad\Longleftrightarrow\qquad
 M'+m{\cal H}M=\text{sources}.
 \label{eq:app-cov-damping}
\end{align}

In a homogeneous, isotropically expanding flow with no sources,
a central moment of velocity order $m$ therefore redshifts as $a^{-m}$. The damping is set by the total velocity order, including traces.

To combine the equations at $m=2$ we add the $(\ell,n)=(0,1)$ and
$(2,0)$ members, the trace and tracefree rank-two equations. This gives us a
single equation for the second moment
$M_{ab}=\langle v_av_b\rangle$, whose trace is $M^{(1)}=3P/\rho_m$ and tracefree part is
$M^{(0)}_{ab}=\pi_{ab}/\rho_m$:
\begin{align}
 \big(\dot M_{ab}\big)_\perp+\frac23\Theta M_{ab}
 +2M_{c(a}\big(\si_{b)}{}^{c}+\om_{b)}{}^{c}\big)
 +\cd^cM_{abc}+M_{abc}\cd^c\ln\rho_m=0.
 \label{eq:app-cov-cumulant2}
\end{align}
Here we have collected the two members at $m=3$ into
$M_{abc}=\langle v_av_bv_c\rangle
=M^{(0)}_{abc}+\tfrac35h_{(ab}M^{(1)}_{c)}$, so both the tracefree third moment and its vector trace enter
the pressure evolution. In its Newtonian limit,
eq.~\eqref{eq:app-cov-cumulant2} is eq.~(36) of \cite{Garny:2022tlk}.

Appendix~\ref{app:integration} gives the scalar system including the third
cumulant, and appendix~\ref{app:convergence} tests its contribution. The relativistic hierarchy and its angular
decomposition are given in \cite{ClarksonMaartens:CovariantCompanion}.

\subsection{Linear scalar perturbations}
\label{sec:vpt-scalars}

We use Poisson gauge,
\begin{align}
 \ud s^2=a^2\left[-(1+2\Phi)\ud\tau^2
 +(1-2\Psi)\delta_{ij}\ud x^i\ud x^j\right],
 \label{eq:main-poisson-metric}
\end{align}
with comoving wavenumber $k$, Fourier convention
$\partial_i\to{\rm i}k_i$ and $\hat k_i=k_i/k$. The lapse potential
$\Phi$ enters the Euler equation and $\Psi$ describes the spatial
curvature. Here $v_i$, $A_i$ and $\si_{ij}$ denote physical spatial components
at linear order, while $\partial_i$ acts on comoving coordinates.
Covariant coordinate components of acceleration and shear are $aA_i$
and $a^2\si_{ij}$, respectively. We write
$\rho_m=\bar\rho_m(1+\delta_m)$, $P=\bar P+\delta P$,
$v_i=\partial_i v$ and $\vartheta=\partial_i v^i$. We denote the
scalar amplitude of the anisotropic pressure by $\pi_{\rm s}$:
\begin{align}
 \pi^i{}_j(\bm k,\tau)
 =\left(\hat k^i\hat k_j-\frac13\delta^i{}_j\right)\pi_{\rm s}(\bm k,\tau).
 \label{eq:main-pi-scalar}
\end{align}
The Fourier divergence ${\rm i}k_j\pi^j{}_i$ is
$2{\rm i}k_i\pi_{\rm s}/3$. Both isotropic and
anisotropic pressure contribute to the longitudinal acceleration. On
subhorizon scales their relevant combination is
\begin{align}
 S(k,a)=\frac{\delta P+2\pi_{\rm s}/3}{\bar\rho_m(\mathcal Hf)^2},
 \label{eq:main-S-mu-Q}
\end{align}
and the momentum conservation (Euler) equation becomes
\begin{align}
 a A_i&=v_i'+\mathcal H v_i+{\rm i}k_i\Phi
 =-{\rm i}k_i(\mathcal Hf)^2S.
 \label{eq:main-A-scalar}
\end{align}
The relativistic projected pressure gradient also contains
$\bar P'v_i$. For the growing modes its contribution is suppressed
by $(\mathcal H/k)^2$ relative to the spatial pressure gradient, as
are the pressure corrections to the Poisson equation. We retain the
leading subhorizon terms; the relativistic equations and this reduction
are given in appendix~\ref{app:relativistic-scalars}.

Rest-mass conservation gives $\delta_m'=-\vartheta$ in this limit.
Combining it with the divergence of eq.~\eqref{eq:main-A-scalar} and
$-k^2\Phi=3\mathcal H^2\Omega_m\delta_m/2$ yields
\begin{align}
 \delta_m''+\mathcal H\delta_m'
 -\frac32\mathcal H^2\Omega_m\delta_m
 +k^2(\mathcal Hf)^2S=0.
 \label{eq:main-VPT-system1}
\end{align}
To evolve the pressure, we perturb about a spatially flat background carrying a homogeneous
dispersion, so that $\bar M_{ab}=(\bar P/\bar\rho_m)\delta_{ab}$.
The anisotropic pressure, acceleration, shear, vorticity and third
moment vanish in the background. Density, expansion and isotropic
dispersion each have a background value and a perturbation. At this order $\cd_a\to{\rm i}k_a/a$ and $\dot X=X'/a$.
For the scalar velocity $v_a=({\rm i}kv)\hat k_a$, the expansion
perturbation and shear are
\begin{align}
 \delta\Theta\simeq\cd^av_a=\frac{\vartheta}{a},
 \qquad
 \si_{ab}=\cd_{\langle a}v_{b\rangle}
 =\frac{\vartheta}{a}\,\hat k_{\langle a}\hat k_{b\rangle}.
 \label{eq:app-cov-kinematics}
\end{align}
The expansion perturbation and shear have the same velocity-divergence
amplitude.

When we linearise eq.~\eqref{eq:app-cov-cumulant2} we therefore pick up
an isotropic source through $\delta\Theta$ and an anisotropic one
through $\si_{ab}$, and by eq.~\eqref{eq:app-cov-kinematics} their sum is
purely longitudinal:
\begin{align}
 \delta M_{ab}'+2{\cal H}\,\delta M_{ab}
 =-2\,\frac{\bar P}{\bar\rho_m}\,\vartheta\,\hat k_a\hat k_b
 -a\,\cd^c\delta M_{abc}.
 \label{eq:app-cov-tau2-lin}
\end{align}
The pressure has its own evolution equation and is coupled to the density.
We write the scalar dispersion perturbation as
\begin{align}
 \delta C_{ij}=(\mathcal Hf)^2
 \big(\delta\epsilon\,\delta_{ij}+g\,\hat k_i\hat k_j\big),
 \qquad S=\epsilon\delta_m+\delta\epsilon+g.
 \label{eq:main-pressure-decomposition}
\end{align}
The decomposition agrees with eq.~(58) of \cite{Garny:2022tlk}.
The term $\epsilon\delta_m$ arises because the pressure is density
weighted. With $\mu=3\Omega_m/(2f^2)$, the linear second-moment
equations at $c_{\max}=2$ are
\begin{align}
 (\delta\epsilon)'+2\mathcal Hf(\mu-1)\delta\epsilon&=0,
 \notag\\
 g'+2\mathcal Hf(\mu-1)g&=2\epsilon\delta_m'.
 \label{eq:main-dispersion-scalars}
\end{align}
Here $g$ multiplies the full longitudinal tensor $\hat k_i\hat k_j$,
including its trace. The source in eq.~\eqref{eq:main-dispersion-scalars}
therefore generates $g$ alone. The independent isotropic mode
$\delta\epsilon$ is unsourced and is set to zero in the growing solution.
These are the two scalar equations in eq.~(64) of \cite{Garny:2022tlk}.
Differentiating eq.~\eqref{eq:main-pressure-decomposition} then gives
\begin{align}
 S'+2\mathcal Hf(\mu-1)S
 &=3\epsilon\delta_m'+\mathcal Hf Q_\epsilon\delta_m,
 \label{eq:main-VPT-system}\\
 Q_\epsilon&=\frac{\epsilon'}{\mathcal Hf}+2(\mu-1)\epsilon.
 \label{eq:main-mean-source}
\end{align}
Of the coefficient $3$ multiplying $\epsilon\delta_m'$, one comes
from differentiating the density-weighting term $\epsilon\delta_m$
and two from the evolution of the longitudinal dispersion $g$.
The term $Q_\epsilon$
accounts for the specified time dependence of the mean dispersion.
In VPT it is supplied by the ensemble-averaged hierarchy, including
advection, pressure--velocity correlations and, when retained,
third-cumulant correlations \cite{Garny:2022tlk}. For freely redshifting dispersion,
$(\mathcal Hf)^2\epsilon\propto a^{-2}$ and $Q_\epsilon=0$.
The prescribed history represents production of random motions by
nonlinear structure formation through a nonzero $Q_\epsilon$, with
$Q_\epsilon/\epsilon\simeq4.8$ today.
A long-wavelength density perturbation can change the abundance and
evolution of the structures producing those motions. Here we prescribe
their mean production and omit its spatial modulation; density,
velocity and dispersion perturbations still evolve through the coupled
equations above. The resulting mean is maintained by production,
whereas source-free homogeneous pressure in
eqs.~\eqref{eq:vpt-cov-P}--\eqref{eq:vpt-cov-pi} redshifts freely.
Section~\ref{sec:nonlinear-scope} discusses the omitted modulation and
appendix~\ref{app:production-modulation} quantifies its possible effect
in an illustrative model.

\begin{figure}[t]
\centering
\includegraphics[width=\textwidth]{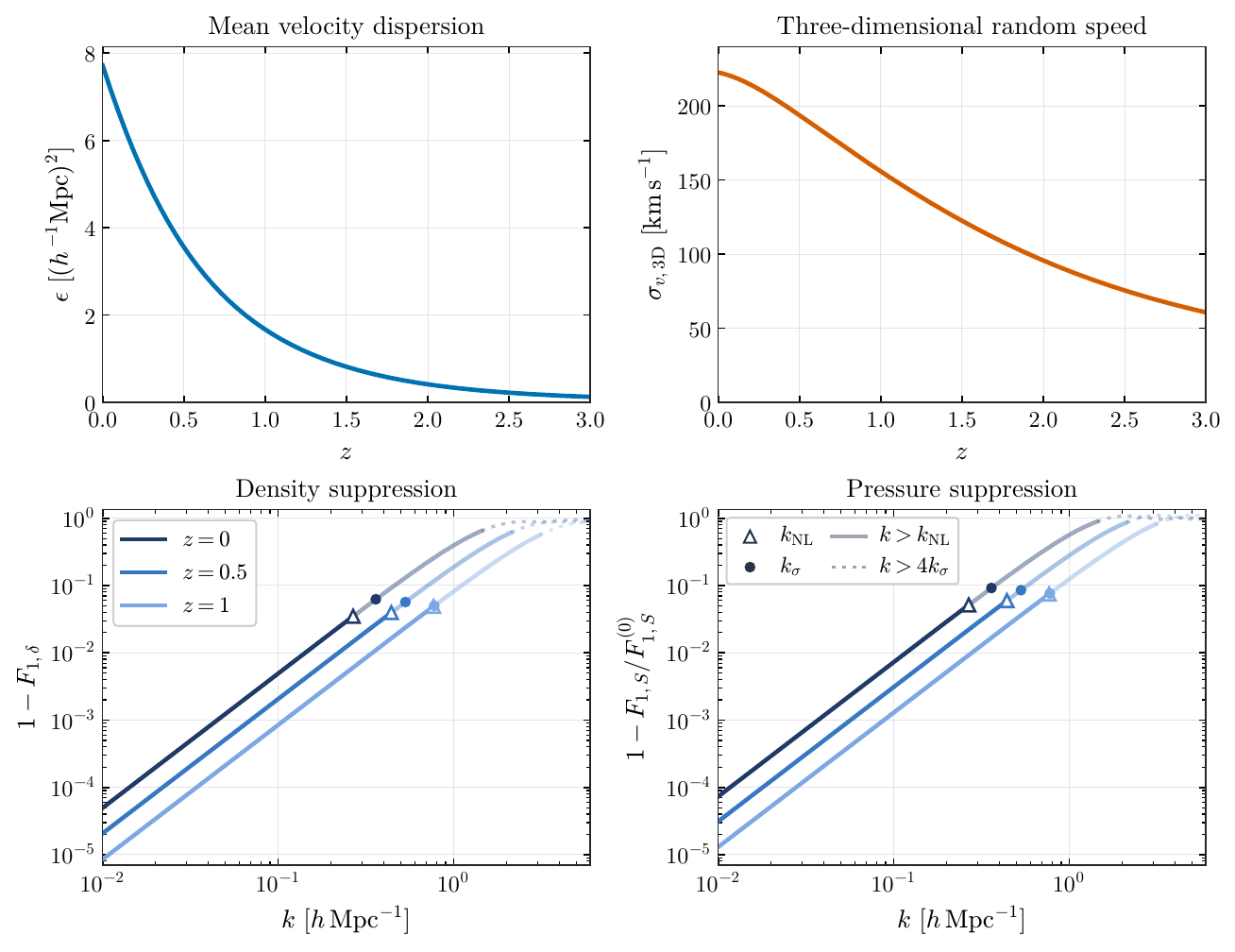}
\caption{The prescribed mean dispersion and the linear $c_{\max}=2$ VPT
kernels. Top: $\epsilon(z)$ and the three-dimensional rms random
particle speed $\sigma_{v,\mathrm{3D}}=\mathcal Hf\sqrt{3\epsilon}$ about
the local mean, including internal halo motions. Bottom: $1-F_{1,\delta}$ and
$1-F_{1,S}/F_{1,S}^{(0)}$, where
$F_{1,S}^{(0)}=\lim_{k\to0}F_{1,S}$. Their leading large-scale
departures are proportional to $k^2\epsilon$. Triangles mark
$k_{\rm NL}$ and circles mark $k_\sigma$. Pale curves extend beyond
$k_{\rm NL}$; dotted segments extend beyond $4k_\sigma$, where the
finite truncation can generate oscillations and sign changes.}
\label{fig:vpt-epsilon-kernels}
\end{figure}

Once $\epsilon(a)$ and initial conditions selecting the growing mode
are specified, the coupled density and pressure equations form a closed
linear system. At early times, when $k^2\epsilon\ll1$, we select the
dust growing mode and its associated pressure perturbation
(eq.~\eqref{eq:app-VPT-initial}).
We solve for the spatial fluctuations retaining the full dependence
on $k^2\epsilon$, without expanding in this parameter.
We define the kernels by
\begin{align}
 \delta_m(\bm k,a)&=D(a)F_{1,\delta}(k,a)\delta_{L0}(\bm k),
 \label{eq:main-VPT-kernels1}\\
 S(\bm k,a)&=D(a)F_{1,S}(k,a)\delta_{L0}(\bm k),
 \label{eq:main-VPT-kernels}
\end{align}
where $\delta_{L0}$ is the pressureless linear density field extrapolated
to today. The factor $D(a)$ is the scale-independent dust growth;
the kernels contain the VPT scale dependence. For $k^2\epsilon\ll1$ the density kernel
approaches unity. Near $k_\sigma$, pressure gradients oppose
gravitational infall and reduce the growing mode. The kernels in
figure~\ref{fig:vpt-epsilon-kernels} describe this collisionless form of
Jeans suppression.

At sufficiently large $k$, the $c_{\max}=2$ solution varies on a
timescale short compared with the expansion and the evolution of
$\epsilon$. Equation~\eqref{eq:main-VPT-system} then reduces to
$S'\simeq3\epsilon\delta_m'$, so its rapidly varying part satisfies
$S\simeq3\epsilon\delta_m$. Neglecting the expansion and gravitational
terms in eq.~\eqref{eq:main-VPT-system1} gives
\begin{align}
 \delta_m''+3k^2(\mathcal Hf)^2\epsilon\delta_m\simeq0.
 \label{eq:main-high-k}
\end{align}
In this limit the truncated equation has oscillatory solutions.
Following the relative motion and phase mixing of particle streams
requires the higher members of the velocity hierarchy
\cite{Garny:2022tlk,Garny:2025ovs}. Nonlinear density evolution
becomes important first: defining $k_{\rm NL}$ by
$\Delta_{m,{\rm lin}}^2(k_{\rm NL},z)=1$, we find
$k_{\rm NL}/k_\sigma=0.74$, $0.83$ and $0.99$ at $z=0$, $0.5$ and $1$.
The high-$k$ continuations illustrate the linear equations
outside their predictive range.

We use $c_{\max}=2$ for the acceleration spectrum and spatial and radial
windows, and $c_{\max}=3$ for every local Hubble and deceleration multipole
in section~\ref{sec:observer-multipoles}. The latter involve pressure time
derivatives, whose evolution contains the third cumulant explicitly. Both
calculations set the mean fourth cumulant to zero.
Appendix~\ref{app:convergence} compares the truncations, and
appendix~\ref{app:windows} identifies the short-wavelength contributions
to each windowed variance.

\Needspace{11\baselineskip}
\subsection{Acceleration spectrum}
\label{sec:vpt-acceleration-kernel}
\label{sec:vpt-transfer}

We compare VPT with dust at a fixed present-day density field, so that
both have the same $\sigma_8$ today. Define
\begin{align}
 \delta_0(\bm k)&\equiv\delta_m(\bm k,1),
 \qquad \delta_m(\bm k,a)=\mathcal D(k,a)\delta_0(\bm k),
 \notag\\*
 \mathcal D(k,a)&=D(a)\frac{F_{1,\delta}(k,a)}{F_{1,\delta}(k,1)},
 \qquad \mathcal D(k,1)=1.
 \label{eq:main-DV-definition}
\end{align}
Here $\mathcal D(k,a)$ is the scale-dependent VPT growth, while
$D(a)$ remains the pressureless growth factor. The differences then
measure changes in velocity and pressure for the
same final density field. A comparison at fixed primordial amplitude
would change the final density spectrum by $F_{1,\delta}^2(k,1)$
and give the two models different present-day $\sigma_8$.

Using eq.~\eqref{eq:main-A-scalar}, we define the acceleration
coefficient by
\begin{align}
 a A_i=-{\rm i}k_iC_A(k,a)\delta_m,
 \qquad C_A=(\mathcal Hf)^2\frac{F_{1,S}}{F_{1,\delta}}.
 \label{eq:main-CA-definition}
\end{align}
The density equation supplies the consistency identity,
\begin{align}
 C_A(k,a)=\frac{1}{k^2\mathcal D}
 \left[\frac32\mathcal H^2\Omega_m\mathcal D
 -\mathcal D''-\mathcal H\mathcal D'\right].
 \label{eq:main-CA-growth}
\end{align}
For pressureless evolution $F_{1,\delta}=1$, $F_{1,S}=0$ and $C_A=0$.
At long wavelengths $C_A$ multiplies a density gradient in the same
way as an effective sound-speed coefficient. Section~\ref{sec:nonlinear-scope}
and appendix~\ref{sec:cg_dust} compare it with the acceleration obtained by
spatially coarse-graining dust \cite{Baumann:2010tm,Carrasco:2012cv}.

Restoring $c$, the dimensionless acceleration is $d_i=-A_i/(cH)$,
with $A_i$ now expressed as a physical acceleration. Its transfer
from $\delta_0$ is
\begin{align}
 d_i(\bm k,a)&={\rm i}\hat k_iT_A(k,a)\delta_0(\bm k),
 \qquad
 T_A(k,a)=\frac{k}{c\mathcal H(a)}C_A(k,a)\mathcal D(k,a).
 \label{eq:main-d-transfer}
\end{align}
The factor $1/c$ makes $T_A$ dimensionless when $C_A$ is expressed in
$({\rm km}\,{\rm s}^{-1})^2$. We define the spectrum by
\begin{align}
 \big\langle\delta_0(\bm k)\delta_0(\bm k')\big\rangle_{\rm ens}
 &=(2\pi)^3\delta_D(\bm k+\bm k')P_0(k),
 \qquad \Delta_0^2(k)=\frac{k^3}{2\pi^2}P_0(k).
 \label{eq:main-P0}
\end{align}
We use the baryon-corrected Eisenstein--Hu no-wiggle transfer function
\cite{Eisenstein:1997ik}, normalised to $\sigma_8=0.8$ for the cosmology
specified above (numerical details are in appendix~\ref{app:vpt-numerics}).

\begin{figure}[t]
\centering
\includegraphics[width=\textwidth]{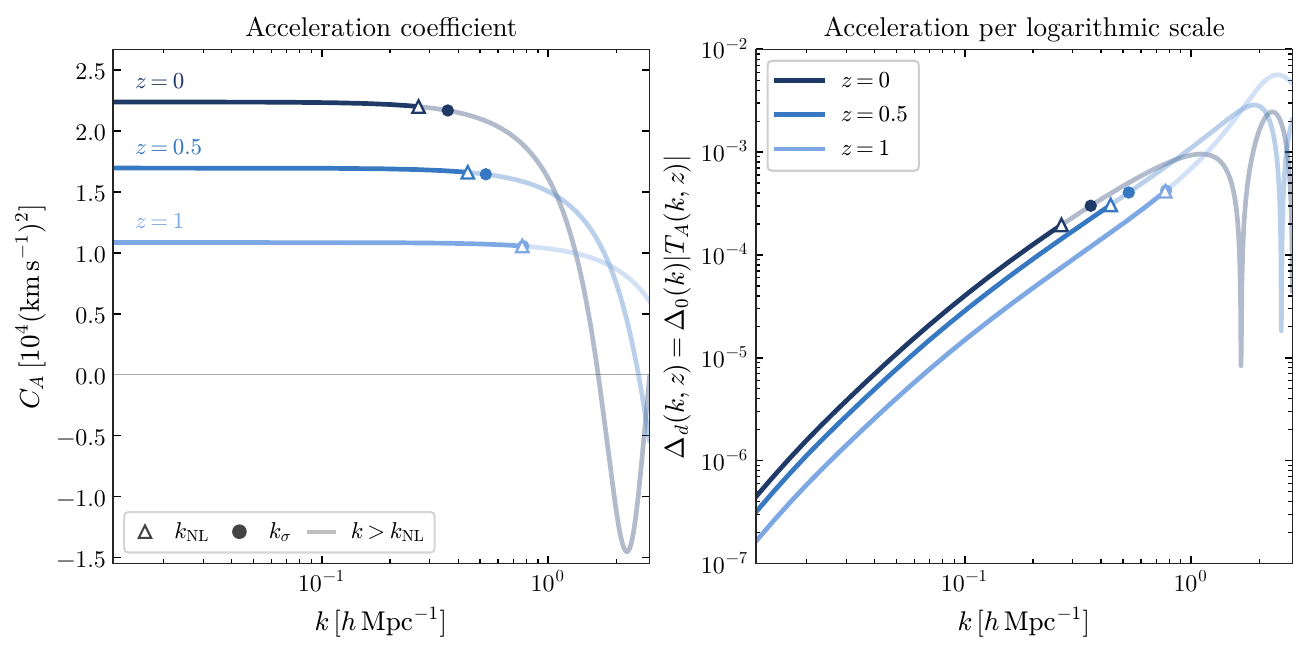}
\caption{Linear $c_{\max}=2$ VPT acceleration with the
prescribed mean dispersion. Left: $C_A(k,z)$ at $z=0,0.5,1$. Right: the dimensionless
acceleration power amplitude $\Delta_d(k,z)$. Triangles mark
$k_{\rm NL}$ and circles mark $k_\sigma$; pale curves begin at
$k_{\rm NL}$. The continuation requires nonlinear density evolution,
and the finite cumulant truncation develops spurious oscillations and
sign changes at still larger wavenumbers. The plotted $\Delta_d$ is the
square root of the variance of $A/(cH)$ per unit $\ln k$. Values of order
$10^{-3}$ occur in the pale continuation outside the linear density regime.}
\label{fig:vpt-acceleration-window}
\end{figure}

The coefficient $C_A$ gives the acceleration produced by a specified
density gradient, but does not determine which scales dominate its
variance. The contribution per logarithmic wavenumber interval is
\begin{align}
 \Delta_d^2(k,z)=\Delta_0^2(k)|T_A(k,z)|^2.
 \label{eq:main-acceleration-dimensionless-power}
\end{align}
As shown in figure~\ref{fig:vpt-acceleration-window}, appreciable power
comes from modes near and above $k_{\rm NL}$ and $k_\sigma$. An
unsmoothed amplitude would therefore depend on modes that this linear
calculation does not control. We quote acceleration amplitudes only
after specifying the window that selects the relevant scales.

\Needspace{13\baselineskip}
\section{Local expansion multipoles}
\label{sec:observer-multipoles}
\subsection{Hubble dipole and quadrupole}

We calculate these local cosmographic coefficients for the matter flow
resolved on a specified spatial scale. Smoothing suppresses the
short-wavelength contributions to the linear variance discussed in
section~\ref{sec:vpt-acceleration-kernel}. We use a Gaussian window,
$W_{\rm G}(kR)=\exp[-(kR)^2/2]$, and compare multipoles of different
angular rank through their sky rms. At a given observer, this measures
the variation of a multipole across the sky; its amplitude also varies
between observers. We denote the VPT--dust difference at the same
present-day density field by $\Delta\eta=\eta_{\rm VPT}-\eta_{\rm dust}$.
For a rank-$\ell$ PSTF contribution
$\Delta\eta_{{\cal A}_\ell}n^{a_1}\cdots n^{a_\ell}$,
angular orthogonality gives \cite{GebbieEllis:2000}
\begin{align}
 \left\langle
  \big(\Delta\eta_{{\cal A}_\ell}n^{a_1}\cdots n^{a_\ell}\big)^2
 \right\rangle_{\bm n}
 =\frac{\ell!}{(2\ell+1)!!}
  \Delta\eta_{{\cal A}_\ell}\Delta\eta^{{\cal A}_\ell}.
 \label{eq:observer-sky-rms}
\end{align}
Unless stated otherwise, we quote the square root of the ensemble
average of this mean square over observers. Section~\ref{sec:observer-variation}
gives the distribution of amplitudes among individual observers.
For the leading acceleration,
\begin{align}
 \sigma_{\eta_A,1}^2(R)
 =\frac{1}{3\ln^2 10}\int_{k_{\min}}^{k_{\max}}\ud\ln k\,
 \Delta_d^2(k,0)W_{\rm G}^2(kR).
 \label{eq:main-acceleration-sky-variance}
\end{align}
The integration limits and numerical details are given in
appendix~\ref{app:vpt-numerics}.

We use $R=30\,h^{-1}{\rm Mpc}$, where the direct contribution from
nonlinear modes is negligible, and a finer radius
$R=13.4\,h^{-1}{\rm Mpc}$. The latter matches the long-wavelength
second moment of a real-space top-hat extending to $z=0.01$
(appendix~\ref{app:numerical-amplitudes}). The matched windows retain
different short-wavelength power, and a catalogue's effective resolution also depends on its
selection and estimator \cite{Adamek:2024Cosmography}.

The shear follows from the velocity divergence. With
$\vartheta=-\delta_m'$ and
$\si_{ij}=a^{-1}\vartheta\hat k_{\langle i}\hat k_{j\rangle}$,
\begin{align}
 \frac{\si_{ij}}{H}
 =-\frac{\partial\mathcal D}{\partial\ln a}
 \hat k_{\langle i}\hat k_{j\rangle}\delta_0.
 \label{eq:observer-shear-transfer}
\end{align}
Define the change in velocity growth by
$\Delta G=\partial\mathcal D/\partial\ln a-fD$. The corresponding
changes in the Hubble multipoles are
\begin{align}
 \frac{\Delta\mathbb H_{ij}}{H}
 &=-\Delta G\,\hat k_{\langle i}\hat k_{j\rangle}\delta_0,
 &
 \frac{\Delta\mathbb H_i}{H}
 &={\rm i}\hat k_iT_A\delta_0.
 \label{eq:observer-H12-transfers}
\end{align}
There is no Hubble octupole. The shear correction changes an existing dust multipole, and the
acceleration supplies a new matter-frame dipole.

At these radii the Hubble-shear correction has sky rms
$2.3\times10^{-5}$ and $2.2\times10^{-4}$, or $0.3\%$ and $1.3\%$
of the dust shear rms. It is the largest correction at zeroth order in
redshift. The acceleration dipole has sky rms $8.4\times10^{-7}$ and
$4.0\times10^{-6}$. The detailed amplitudes and truncation test are in
table~\ref{tab:observer-amplitudes}.

\subsection{Deceleration multipoles}
\label{sec:deceleration-multipoles}

The next term in the distance expansion brings in derivatives of the
local flow. Expanding eq.~\eqref{eq:main-distance-expansion} gives
\begin{align}
 (\ln10)\eta(z,n^a)
 =\left[-\frac{A_a}{H}n^a+\frac{\si_{ab}}{H}n^an^b
 +\frac z2\big(\mathbb Q(n^a)-\langle\mathbb Q\rangle_{\bm n}\big)
 \right]_{\rm o}+O(z^2,\delta_m^2).
 \label{eq:app-observer-eta}
\end{align}
The linear deceleration multipoles follow from Paper I
\cite{ClarksonMaartens:CovariantCompanion} and \cite{Heinesen:2020bej}. At linear scalar order the shear constraint gives
$\cd^b\si_{ab}=2\cd_a\Theta/3$. The deceleration dipole is then
\begin{align}
 \mathbb Q_a
 =\frac{3}{5H^2}\cd_a\Theta
 +\frac{1}{H^2}\left[
 \dot A_{\langle a\rangle}+2(1+q)HA_a\right],
 \label{eq:app-observer-Q}
\end{align}
where $q=-1-\dot H/H^2$ is the background deceleration parameter.
The expansion-gradient term is also present for dust. Velocity
dispersion changes it by altering the growth of density and velocity
perturbations. We label this change \emph{grow}; terms containing $A_a$
or its derivatives explicitly are labelled $A$. Both arise from the same
pressure dynamics and must be added with their relative signs. Writing $d_a=-A_a/H$
in units $c=1$, the acceleration contribution through this order is
\begin{align}
 \eta_A(z,n^a)=\frac{n^a}{\ln10}
 \left[d_a-\frac z2\left((1+q)d_a+
 \frac{\dot d_{\langle a\rangle}}{H}\right)\right]_{\rm o}
 +O(z^2,\delta_m^2).
 \label{etaA}
\end{align}
Here $\dot A_{\langle a\rangle}=h_a{}^b u^c\nabla_cA_b$, and similarly for $d_a$.
The projected time derivative includes changes in both the magnitude
and direction of the acceleration.

The quadrupole and octupole are
\begingroup
\allowdisplaybreaks[0]
\begin{align}
 \mathbb Q_{ij}
 &=-\frac{\dot\si_{ij}}{H^2}
   -2(1+q)\frac{\si_{ij}}{H}
   -\frac{\cd_{\langle i}A_{j\rangle}}{H^2},
 \notag\\
 \mathbb Q_{ijk}
 &=\frac{\cd_{\langle i}\si_{jk\rangle}}{H^2}.
 \label{eq:observer-Q23-kinematic}
\end{align}
\endgroup
These expressions use shear propagation to replace the electric Weyl
tensor. The octupole has the same kinematic form as for dust and gains
its Vlasov correction through the changed shear. The contributions
to the dipole and octupole from changes in growth have the same
velocity-growth source. Their scalar transfers are proportional, although their
different angular ranks give different sky amplitudes and make their
components uncorrelated at the same observer in an isotropic ensemble.
The transfers and angular normalisations are collected in
appendix~\ref{app:observer-normalisation}.

\begin{figure}[t]
\centering
\includegraphics[width=\textwidth]{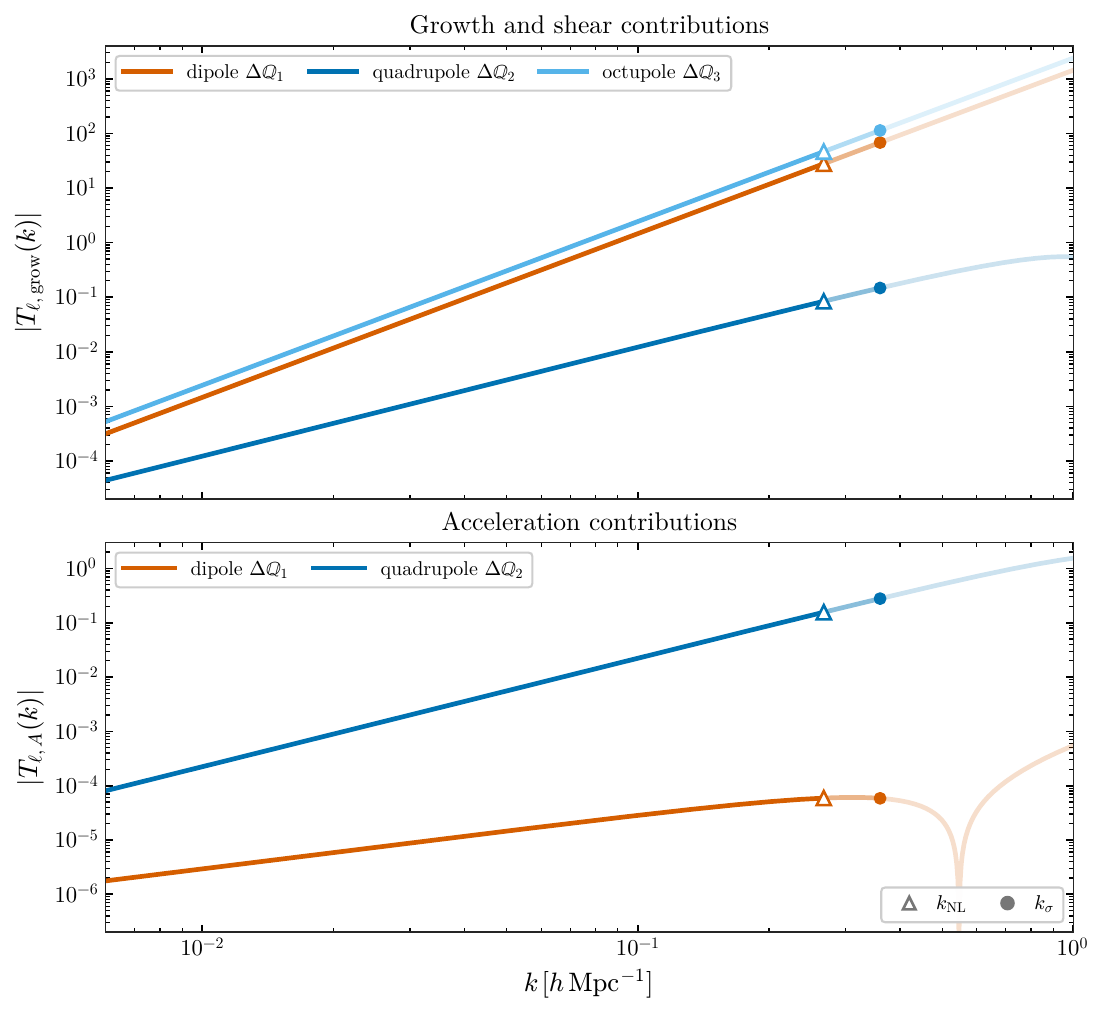}
\caption{VPT--dust corrections to the deceleration multipoles at $z=0$.
The upper panel shows changes in the ordinary growth and shear terms;
the lower panel shows terms containing acceleration explicitly. There
is no explicit acceleration term in the octupole at linear scalar
order. Pale curves begin at $k_{\rm NL}$ and become lighter beyond
$k_\sigma$, marked by triangles and circles respectively.}
\label{fig:vpt-Q-transfers}
\end{figure}

To compare the leading and first redshift-dependent terms at a
representative depth, we average the local Taylor expression with
the volume weight of a cumulative sample ending at $z_{\max}=0.01$.
This replaces $z$ by $\bar z_w=7.5\times10^{-3}$. The growth
contribution to the dipole has sky rms $1.7\times10^{-5}$ at $30\,h^{-1}{\rm Mpc}$ and
$3.6\times10^{-4}$ at $R=13.4\,h^{-1}{\rm Mpc}$. The acceleration transfer evolves slowly today, so its additional
$O(z)$ contribution is suppressed by approximately
$(\bar z_w/2)(1+q_0)$ relative to the intrinsic dipole
(eq.~\eqref{etaA}). The quadrupole
is dominated by the Hubble shear, with a small partial cancellation
from deceleration.

Successive redshift orders can introduce further spatial gradients,
bringing factors of $ck/H_0$. At comoving distance $\chi$, the
combination $k\chi$ compares the distance traversed with the spatial
variation scale $k^{-1}$ of a mode. The spatial Taylor expansion
requires $k\chi\ll1$ for the modes that dominate the signal.
A small value of $z$ alone does not ensure this, since $ck/H_0$
can be large.
Our variance-weighted estimates give $k\chi$ of order unity or larger
at both radii, for dust as well as Vlasov matter
(appendix~\ref{app:numerical-amplitudes}). The local coefficients are
well filtered against nonlinear modes, but extrapolating them across
the sample introduces spatial variations that the first redshift
orders cannot describe reliably. Comparison with a finite-distance fit
requires the light-cone calculation discussed in
section~\ref{sec:model-limits}.
Numerical and exact-spacetime studies likewise show why local
derivatives and coefficients fitted across a finite region must be
distinguished
\cite{MacphersonHeinesen:2025Dipole,Sarma:2025LocalStructures,
HillsHeinesen:2026Szekeres}.

At fixed smoothing radius, $k\chi$ decreases towards the observer.
The CF4 cut at $z\simeq0.01$ addresses a different approximation:
representing very nearby galaxy velocities by a smooth flow
\cite{Kalbouneh:2025jnp}. The linear Vlasov corrections found here
provide no explanation for the deterioration of that fit. We interpret
the low-redshift sensitivity primarily as a limit of this smooth-flow
cosmographic description.

\subsection{Variation between observers}
\label{sec:observer-variation}

\begin{figure}[t]
\centering
\includegraphics[width=\textwidth]{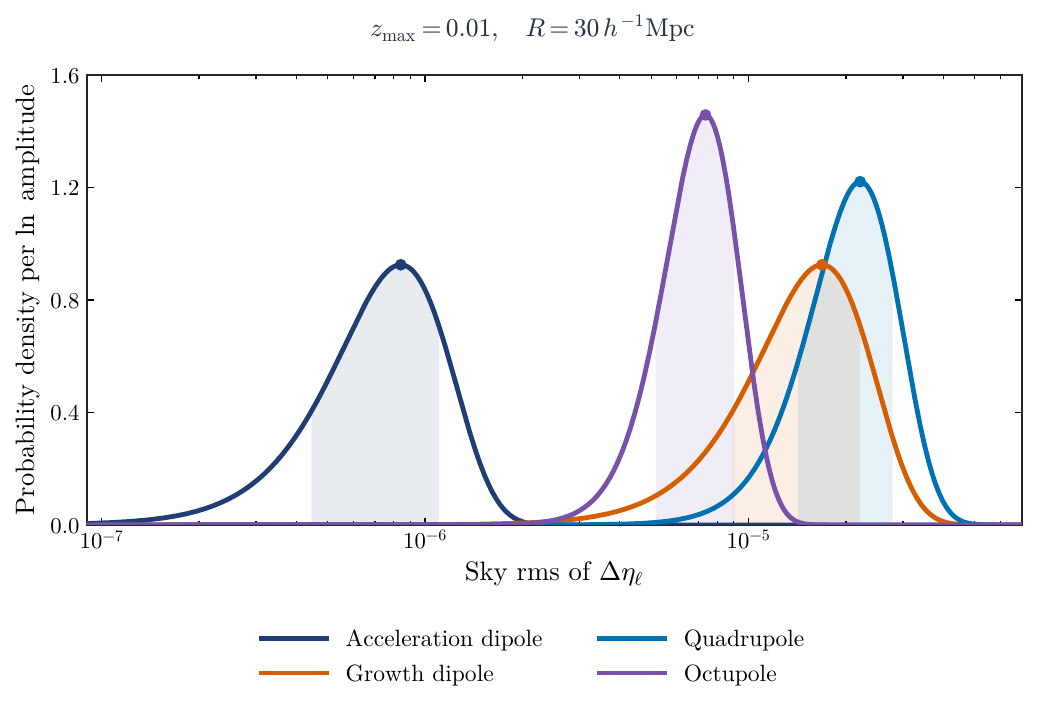}
\caption{Probability distributions of VPT--dust corrections at a random
observer, with $z_{\max}=0.01$ and Gaussian smoothing at
$R=30\,h^{-1}{\rm Mpc}$. The horizontal axis is the sky rms of each
contribution to $\eta$, so different multipole ranks can be compared
directly. The density is per logarithmic amplitude interval; shading
marks the central $68\%$ interval and dots mark the ensemble rms.
The acceleration and growth dipoles are shown separately; the acceleration curve includes its small
$O(z)$ correction. The quadrupole includes Hubble shear and deceleration.}
\label{fig:vpt-observer-multipoles}
\end{figure}

For Gaussian initial conditions, the components of a linear multipole
in an orthonormal PSTF basis are Gaussian. Statistical isotropy gives
them equal variance, so the norm follows a chi distribution with
$2\ell+1$ degrees of freedom. Figure~\ref{fig:vpt-observer-multipoles}
shows the resulting variation between random observers. The
quadrupole includes its leading Hubble-shear term, while the octupole
begins at $O(z)$; sky rms puts them on a common angular normalisation.

Comparing two contributions at the same observer uses their joint
distribution. The acceleration and growth dipoles have correlation coefficient
$0.83$ for the $30\,h^{-1}{\rm Mpc}$ window. Their joint covariance
allows us to ask whether the intrinsic acceleration could dominate
the change in the ordinary growth dipole. This happens for only
$3.7\times10^{-5}$ of random observers in the model
(appendix~\ref{app:observer-probability}), predominantly
where the growth dipole happens to be small. The acceleration's $99$th-percentile sky amplitude is $1.6\times10^{-6}$.
For comparison, the measured CF4a dipole has sky rms $1.26\times10^{-2}$
(section~\ref{sec:intrinsic-dipole}). These probabilities describe random
positions; observers in haloes or Local Group-like environments require a
distribution selected for those environments.

\section{Implications for dust cosmography}
\label{sec:robustness}
\subsection{The intrinsic dipole}
\label{sec:intrinsic-dipole}

Kinetic pressure changes the cosmographic expansion in two ways. It alters
the growth of density and velocity perturbations, changing the shear
and expansion gradients that already enter dust cosmography. It also
accelerates the mean matter flow, producing an intrinsic Hubble
dipole. The latter is absent for dust in its rest frame and is the
most direct signature of the additional matter dynamics.

The measured CMB-frame CF4a dipole in \cite{Kalbouneh:2025jnp}, for
$0.01<z<0.03$, has Legendre coefficient
$\widetilde\eta_1=(2.19\pm0.13)\times10^{-2}$ (their table~3).
For a dipole proportional to $P_1(\cos\theta)$, the sky rms is
\begin{align}
 \frac{|\widetilde\eta_1|}{\sqrt3}
 =(1.26\pm0.08)\times10^{-2}.
 \label{eq:observed-dipole-rms}
\end{align}
Motion of the local matter frame relative to the CMB produces a Doppler
contribution proportional to $1/z$ at lowest order, which dominates near
the observer. The intrinsic acceleration dipole calculated here is
defined in the matter frame and supplies a constant term at that order.
At larger redshifts, $\mathbb Q_1$ and higher cosmographic coefficients
also contribute appreciably to the observed dipole \cite{Kalbouneh:2025jnp}.
The measured dipole is three to four orders of magnitude larger than
our fiducial intrinsic dipole. Even the largest intrinsic amplitude
across the tested dispersion histories remains more than a thousand
times smaller.

The cosmographic fit in eq.~(6.2) of \cite{Kalbouneh:2025jnp} includes
the shear and deceleration anisotropies but fixes the intrinsic Hubble
dipole to zero. Pressure changes coefficients already present in that
fit, as well as generating the additional dipole.

The growth correction is substantially larger than the acceleration
dipole. At the finer radius, the local correction to $\mathbb Q_1$
has an rms component $0.22$ along a fixed axis, about half the fitted
uncertainty of $0.4$ in \cite{Kalbouneh:2025jnp}. This gives the scale
of a possible correction to the fit. Predicting its fitted value
requires the finite-distance calculation, with the redshift weights
and common multipole axis used in that analysis.

The acceleration-dipole test proposed in \cite{ClarksonBarrett:1999}
can be implemented by allowing $\mathbb H_1/\mathbb H_0$ to vary
alongside $v_{\rm o}$ and $\mathbb Q_1$ over $0.01<z<0.1$, using the full
boosted distance model. Their distinct constant, $1/z$ and $O(z)$ dipole
contributions could constrain an intrinsic dipole and test whether fixing
$\mathbb H_1=0$ biases the inferred velocity or deceleration dipole.
Through $\mathbb H_a=-A_a$, this would give an empirical constraint on
the acceleration of the fitted flow. Restricting all multipoles to a common
axis constrains the acceleration component along that axis; a bound on
its full magnitude would also require fitting its direction. Interpreting
the result as a matter-frame acceleration then depends on the tracer
relation in section~\ref{sec:galaxy-tracers}.

\subsection{Velocity dispersion and nonlinear evolution}
\label{sec:nonlinear-scope}

The intrinsic dipole remains far below the measured CF4a dipole throughout
our tested range of mean dispersions. Its amplitude scales approximately
with the mean dispersion, while including the third cumulant changes the
observer amplitudes by less than one percent
(table~\ref{tab:observer-amplitudes}). These variations leave the hierarchy
of corrections found in linear theory intact.

Nonlinear structure can change the pressure perturbations even at fixed mean
dispersion. A long density mode changes the abundance and internal motions
of the structures producing random velocities, and hence the dispersion
in that region. Our example in appendix~\ref{app:production-modulation}
shows that this modulation can change the acceleration dipole by order
unity. Calculating it requires the nonlinear pressure--density and
pressure--velocity correlations. Smoothing the linear fields removes
short-scale fluctuations but cannot recover these correlations.

The Vlasov pressure describes random particle velocities about the mean
flow at each point. Spatial averaging also produces effective pressure
from variations of that flow and unresolved gravitational fields
\cite{Baumann:2010tm,Carrasco:2012cv,Nascimento:2024rbv}.
Appendix~\ref{sec:cg_dust} compares the coefficients relating these
accelerations to the same long-wavelength density gradient. With linear
short modes above
$\Lambda=k_{\rm NL}(0)\simeq0.27\,h\,{\rm Mpc}^{-1}$, the perturbative
coarse-grained coefficient agrees with our fiducial present-day VPT value
to about $5\%$. In virialised objects, kinetic and gravitational contributions
can cancel in the slowly varying, long-wavelength acceleration
\cite{Baumann:2010tm}; their relation to the local Vlasov pressure
requires a nonlinear calculation.

\label{sec:model-limits}
Our results describe local Taylor coefficients. Comparison with a
catalogue fit requires distances along the light cone, including the
observer, source and integrated terms, with the catalogue's angular
selection, redshift weights and fitted multipole directions. The gradient
test in section~\ref{sec:deceleration-multipoles} shows why this is necessary
even at low redshift. Relating the matter flow to the observed galaxies
is a further physical input, which we discuss next.

\subsection{Galaxy tracers}
\label{sec:galaxy-tracers}

A galaxy population can define a different mean flow from the matter
energy frame. Following the relative-velocity notation of Paper I,
we write its 4-velocity as
$u^a_{\rm g}=u^a+\w^a_{\rm g}+O(\w_{\rm g}^2)$, with
$u_a\w^a_{\rm g}=0$. Its acceleration satisfies
\begin{align}
 h^a{}_b A^b_{\rm g}=A^a+\dot\w_{\rm g}^{\langle a\rangle}
 +\left(\frac13\Theta h^{ab}+\si^{ab}+\om^{ab}\right)\w_b^{\rm g}
 +O(\w_{\rm g}^2).
 \label{eq:galaxy-acceleration}
\end{align}
The intrinsic Hubble dipole of this flow is $-A^a_{\rm g}$.
For galaxies whose mean velocity follows the matter energy frame,
$\w^a_{\rm g}=0$, it is the dipole calculated here. For a geodesic
galaxy congruence, the relative velocity instead obeys
$\dot\w_{\rm g}^{\langle a\rangle}+H\w^a_{\rm g}=-A^a$
at linear order about FLRW, and the intrinsic dipole vanishes.

Predicting $\w^a_{\rm g}$ for a selected galaxy population requires a
model for the galaxies' velocity bias and velocity dispersion
\cite{Biagetti:2015hva,Matsubara:2019tyb}. The catalogue selection
then determines which population contributes to the observed mean
flow. Together, the matter dynamics and this tracer model determine
the acceleration that enters a prediction for galaxy distances.

\section{Conclusion}
\label{sec:conclusion}

We have calculated how velocity dispersion changes the local acceleration,
shear and expansion gradients by connecting the covariant equations of
Paper I to linear Vlasov perturbation theory. At zeroth order in redshift,
the largest correction is to the Hubble-shear quadrupole: $0.3\%$ and
$1.3\%$ of the dust shear rms at the two Gaussian radii. The intrinsic
acceleration dipole remains more than a thousand times smaller than the
measured CMB-frame CF4a dipole, which includes the much larger contribution
from motion between the matter and CMB frames.

At the next redshift order, the growth-induced dipole can exceed the shear
correction. Pressure therefore affects the deceleration anisotropies
already present in dust cosmographic fits more strongly than it generates
an intrinsic dipole.

The numerical amplitudes rely on the mean dispersion history estimated
from halo structure and abundance in \cite{Garny:2025zlq}, which we adopt
as an input to the linear VPT equations. Nonlinear evolution can change
the pressure perturbations, and a prediction for a galaxy sample also
depends on how its mean flow follows the matter
(section~\ref{sec:galaxy-tracers}). The VPT calculation assumes an FLRW
background and cannot independently test the cosmological principle;
its numerical predictions also depend on the adopted $\Lambda$CDM cosmology.

At fixed smoothing radius and observer velocity, the intrinsic and $O(z)$
growth terms decrease relative to the Doppler dipole as $z$ and $z^2$,
respectively. The linear Vlasov corrections therefore provide no
explanation for the deterioration of the CF4 fit below $z\simeq0.01$.
We interpret this sensitivity primarily as a limit of the smooth-flow
cosmographic description. The size of $k\chi$ also shows that the local
Taylor expansion lacks a small gradient parameter at our representative
depth, for dust as well as Vlasov matter. Connecting these local
coefficients to a catalogue fit requires a finite-distance calculation.

Covariant cosmography itself does not assume an FLRW background.
CMB isotropy and the Copernican principle do not establish homogeneity
when the fundamental observers accelerate \cite{ClarksonBarrett:1999}.
The intrinsic-dipole test described in section~\ref{sec:intrinsic-dipole}
would measure $\mathbb H_a=-A_a$ for the fitted flow and establish how
fixing $\mathbb H_1=0$ affects the inferred velocity and deceleration
dipole. It would directly test the geodesic-flow assumption used in
arguments for the cosmological principle.

\Needspace{9\baselineskip}
\acknowledgments
We thank Christian Marinoni for helpful comments on the manuscript.
CC is supported by STFC (UK) grant ST/X000931/1. CC also acknowledges
OpenAI for providing complimentary access to ChatGPT and Codex through
the ChatGPT for Academic Researchers programme. The authors used OpenAI ChatGPT Astra 6.0 to assist with algebraic consistency checks, numerical implementation, figure preparation, literature comparisons, and manuscript drafting and editing. All analytical results were independently rederived and all numerical results reproduced using separate code developed with Anthropic’s Claude Opus 5.5. Code to reproduce the results is available on request from the authors. The authors take full responsibility for the calculations, interpretation and final text.

\appendix
\section{Velocity cumulants and relativistic equations}
\label{sec:covariant-to-vpt}
\subsection{Velocity cumulants}
\label{app:velocity-cumulants}

Let ${\sf f}_{\rm ps}(\bm x,\bm p,\tau)$ denote the phase-space
distribution, where $\bm p$ is the comoving momentum per unit mass.
It is normalised by $\int\ud^3\bm p\,{\sf f}_{\rm ps}=1+\delta_m$.
The cumulant generating function is
\begin{align}
 {\exp} [{C(\bm x,\bm l,\tau)}]
 &=\int\ud^3{\bm p}\,{\rm e}^{\bm l\cdot\bm p/a}\,{\sf f}_{\rm ps}(\bm x,\bm p,\tau)\,,
 \notag\\
 C_{i_1\cdots i_c}
 &=\left.\frac{\partial^cC}{\partial l_{i_1}\cdots\partial l_{i_c}}
 \right|_{\bm l=0}.
 \label{eq:main-C-generator}
\end{align}

The zeroth cumulant gives $\ln(1+\delta_m)$, the first gives the bulk
peculiar velocity and the second gives the velocity-dispersion tensor.
Writing the fourth central moment as $M_{ijkl}$ gives
\begin{align}
 M_{ijkl}=C_{ijkl}+C_{ij}C_{kl}+C_{ik}C_{jl}+C_{il}C_{jk}.
 \label{eq:fourth-central-moment}
\end{align}
Thus setting the connected fourth cumulant $C_{ijkl}$ to zero leaves
the products of second moments in the fourth central moment.

For scalar perturbations, each
PSTF moment takes the form
\begin{align}
 X_{{\cal A}_\ell}=X\,\hat k_{\langle{\cal A}_\ell\rangle},
 \qquad
 \hat k_{\langle{\cal A}_\ell\rangle}
 \equiv\hat k_{\langle a_1}\cdots\hat k_{a_\ell\rangle},
\end{align}
and its divergence and projected tracefree gradient are
\begin{align}
 \cd^{a_\ell}X_{{\cal A}_\ell}
 &=\frac{{\rm i}k}{a}\,\frac{\ell}{2\ell-1}\,
 X\,\hat k_{\langle{\cal A}_{\ell-1}\rangle},
 &
 \cd_{\langle a_{\ell+1}}X_{{\cal A}_\ell\rangle}
 &=\frac{{\rm i}k}{a}\,X\,\hat k_{\langle{\cal A}_{\ell+1}\rangle}.
 \label{eq:app-cov-ladder}
\end{align}
Both rules follow from
$\hat k^{a_\ell}\hat k_{\langle{\cal A}_\ell\rangle}
=[\ell/(2\ell-1)]\hat k_{\langle{\cal A}_{\ell-1}\rangle}$.
For $\ell=2$, the divergence gives
$\cd^b\pi_{ab}=2{\rm i}k_a\pi_{\rm s}/(3a)$,
as used in eq.~\eqref{eq:main-A-scalar}.

\subsection{Linear scalar equations in Poisson gauge}
\label{app:relativistic-scalars}

The matter-frame derivative of a scalar with nonzero background is
$\cd_i X=a^{-1}(\partial_i\delta X+\bar X'v_i)$ at linear order.
Our scalar conventions are
\begin{align}
 P=\bar P+\delta P,
 \qquad \rho_m=\bar\rho_m(1+\delta_m),
 \qquad v_i=\partial_iv,
 \qquad \vartheta\equiv\partial_iv^i=\nabla^2v.
 \label{eq:main-scalar-definitions}
\end{align}
\Needspace{6\baselineskip}
The kinematic relations are
\begingroup
\allowdisplaybreaks[0]
\begin{align}
 a A_i&=v_i'+\mathcal H v_i+\partial_i\Phi,
 \notag\\
 a\,\delta\Theta&=\vartheta-3\Psi'-3\mathcal H\Phi,
 \qquad a\si_{ij}=\partial_{\langle i}v_{j\rangle}.
 \label{eq:relativistic-kinematics}
\end{align}
\endgroup
The scalar momentum conservation equation therefore includes the projected
background-pressure derivative,
\begin{align}
 a A_i=-\frac{{\rm i}k_i}{\bar\rho_m}
 \left(\delta P+\frac23\pi_{\rm s}+\bar P'v\right).
 \label{eq:relativistic-A-scalar}
\end{align}
\Needspace{10\baselineskip}
\begin{samepage}
The tracefree spatial Einstein equation fixes the potential difference.
Substituting it into the trace equation gives
\begin{align}
 k^2(\Phi-\Psi)&=8\pi Ga^2\pi_{\rm s},
 \label{eq:main-slip}\\
 \Psi''+2\mathcal H\Psi'+\mathcal H\Phi'
 +(2\mathcal H'+\mathcal H^2)\Phi
 &=4\pi Ga^2\left(\delta P+\frac23\pi_{\rm s}\right).
 \label{eq:main-trace-Einstein}
\end{align}
\par\end{samepage}

These fix the potential and anisotropic-pressure conventions
\cite{Ma:1995ey}; no barotropic relation between $\delta P$ and $\delta_m$
is assumed. Rest-mass conservation and momentum conservation give
\begin{align}
 \delta_m'+\vartheta-3\Psi'&=0,
 \label{eq:main-continuity}\\
 v'+\left(\mathcal H+\frac{\bar P'}{\bar\rho_m}\right)v
 &=-\frac{1}{\bar\rho_m}
 \left(\bar\rho_m\Phi+\delta P+\frac23\pi_{\rm s}\right).
 \label{eq:main-Euler}
\end{align}

Multiplying the Euler equation by $-k^2$ gives
$\vartheta'+(\mathcal H+\bar P'/\bar\rho_m)\vartheta
=k^2[\Phi+(\delta P+2\pi_{\rm s}/3)/\bar\rho_m]$.
We differentiate the continuity equation and substitute this expression
for $\vartheta'$. Using $\vartheta=-\delta_m'+3\Psi'$ in the Hubble
damping term, then subtracting $4\pi Ga^2\bar\rho_m\delta_m$ from both
sides, gives
\begin{align}
 &\delta_m''+\mathcal H\delta_m'-4\pi Ga^2\bar\rho_m\delta_m
 +\frac{k^2}{\bar\rho_m}\left(\delta P+\frac23\pi_{\rm s}\right)
 \notag\\
 &\hspace{12mm}
 =3(\Psi''+\mathcal H\Psi')
 -\left[k^2\Phi+4\pi Ga^2\bar\rho_m\delta_m\right]
 +\frac{\bar P'}{\bar\rho_m}\vartheta.
 \label{eq:main-density-master}
\end{align}

On subhorizon scales, the Poisson relation
$-k^2\Phi=4\pi Ga^2\bar\rho_m\delta_m$ cancels the square bracket.
The remaining metric terms are smaller than the ordinary dust terms by
$(\mathcal H/k)^2$. The pressure-dependent metric corrections and the
projected $\bar P'v$ contribution are also smaller than the spatial
pressure contribution by this factor.\footnote{The ordinary dust metric
terms need not be smaller than the pressure contribution: their ratio
is of order $(\mathcal H/k)^2/(k^2f^2\epsilon)$. A relativistic
VPT--dust comparison must include those terms consistently in both
models, together with the pressure correction to the background.}
Dropping these relativistic corrections and evolving the dispersion
with eq.~\eqref{eq:main-dispersion-scalars} gives the two equations used
in section~\ref{sec:vpt-scalars}, at $c_{\max}=2$:
\begin{align}
 \delta_m''+\mathcal H\delta_m'
 -\frac32\mathcal H^2\Omega_m\delta_m
 &=-k^2(\mathcal Hf)^2S,\notag\\
 S'+2\mathcal Hf(\mu-1)S
 &=3\epsilon\delta_m'+\mathcal Hf Q_\epsilon\delta_m.
 \label{eq:app-final-evolution}
\end{align}
Here $S$ and $Q_\epsilon$ are defined in
eqs.~\eqref{eq:main-S-mu-Q} and \eqref{eq:main-mean-source}.
The pressure gradient opposes gravitational growth, while the evolving
density and velocity generate the pressure perturbation.

\section{Numerical implementation}
\label{app:vpt-numerics}

We take the mean dispersion history from the halo-based fit of
\cite{Garny:2025zlq}, evaluated using the dust growth factor in the
adopted $\Lambda$CDM background. For each wavenumber we then evolve the
density, velocity and pressure perturbations about that mean, including
the third cumulant where specified in section~\ref{sec:vpt-scalars}.
We normalise the solutions to the same present-day density field, obtain
the acceleration and cosmographic coefficients, and integrate their
variances with the chosen windows.

\subsection{The mean dispersion used in the calculation}
\label{app:vpt-epsilon}

Velocity dispersion is a local moment of the phase-space distribution,
as in eq.~\eqref{eq:main-second-cumulant}. Nonlinear collapse and shell
crossing generate random motion about the local mean flow.
Reference~\cite{Garny:2025zlq} estimates the mean dispersion from halo
velocity-dispersion profiles, using Navarro--Frenk--White density profiles
\cite{Navarro:1997NFW} and a halo mass function calibrated to $N$-body
simulations. For their
fiducial halo boundary of $1.5r_{\rm vir}$, the fitted history is
\begin{align}
 \epsilon(z)=k_\sigma(0)^{-2}D(z)^\alpha\frac{G(z)}{G(0)},
 \qquad G(z)=\left[1+
 \left(\frac{D(z_c)}{D(z)}\right)^{p\Delta\alpha}\right]^{-1/p},
 \label{eq:app-epsilon-fit}
\end{align}
with $k_\sigma(0)=0.36\,h\,{\rm Mpc}^{-1}$, $\alpha=3.23$,
$\Delta\alpha=1.30$, $z_c=1.52$ and $p=5$.
The halo estimate underlying this fit is
\cite{Garny:2022tlk,Garny:2025zlq}
\begin{align}
 \epsilon_{\rm halo}
 &=\int_0^\infty\ud m\,\frac{m}{\bar\rho_m}
 \frac{\ud n}{\ud m}\,\bar\epsilon_{\rm h},
 &
 \bar\epsilon_{\rm h}
 &=\frac{1}{V_{\rm vir}}\int_{V_{\rm out}}\ud^3r\,
 \frac{C^{\rm h}_{ii}(\bm r)}{3(\mathcal Hf)^2}.
 \label{eq:app-epsilon-halo}
\end{align}
Here $C^{\rm h}_{ij}$ is the local dispersion tensor inside a halo.
The integral extends over $V_{\rm out}$ to $1.5r_{\rm vir}$, including
the halo outskirts, and is normalised by the virial volume $V_{\rm vir}$.
The factor $(m/\bar\rho_m)(\ud n/\ud m)\ud m$ gives the fraction of
matter in each halo mass interval. We adopt the fitted history in
eq.~\eqref{eq:app-epsilon-fit} for every occurrence of $\epsilon$ and
$Q_\epsilon$ in the perturbation equations. It supplies the mean random
motion generated by nonlinear structure; the coupled equations then
calculate its effect on the density and velocity perturbations.\footnote{The
halo prescription is an estimate of the ensemble mean. Measurements of the
spatial dispersion field \cite{Buehlmann:2019VelocityDispersion} provide
an independent way to calibrate it.}

At linear order we set
$\bar P/\bar\rho_m=(\mathcal Hf)^2\epsilon$, giving
$\bar P/(\bar\rho_m c^2)\simeq1.8\times10^{-7}$ today and allowing
us to neglect pressure in the background expansion. In the nonlinear
field, $\langle P\rangle/\bar\rho_m$ also includes the correlation
$\langle\delta_m P/\rho_m\rangle$; the fitted mean dispersion alone
does not fix it.

\Needspace{12\baselineskip}
\subsection{Evolution equations and initial conditions}
\label{app:integration}

We use $N=\ln D$ to write the evolution compactly, with
$\ud N/\ud\tau=\mathcal Hf$, and define
$\theta=\ud\delta_m/\ud N=-\vartheta/(\mathcal Hf)$.
This is the normalised velocity divergence, distinct from the covariant
expansion $\Theta$. The pressure variable is
$S=\epsilon\delta_m+g+\delta\epsilon$ from
section~\ref{sec:vpt-scalars}. At $c_{\max}=3$ its evolution also contains the longitudinal divergence
of the third cumulant,
\begin{align}
 J=\frac{{\rm i}k^l\hat k^i\hat k^j C_{ijl}}{(\mathcal Hf)^3}.
 \label{eq:app-third-divergence}
\end{align}
This term transports dispersion between neighbouring points. With the
mean fourth cumulant set to zero, the system we integrate is
\cite{Garny:2022tlk}
\begingroup
\allowdisplaybreaks[0]
\begin{align}
 \frac{\ud\delta_m}{\ud N}&=\theta,\notag\\
 \frac{\ud\theta}{\ud N}+(\mu-1)\theta
 &=\mu\delta_m-k^2S,\notag\\
 \frac{\ud S}{\ud N}+2(\mu-1)S
 &=3\epsilon\theta+Q_\epsilon\delta_m-J,\notag\\
 \frac{\ud J}{\ud N}+3(\mu-1)J
 &=3k^2\epsilon(S-\epsilon\delta_m).
 \label{eq:app-cmax3-system}
\end{align}
\endgroup
The coefficients $\mu$ and $Q_\epsilon$ are defined in
section~\ref{sec:vpt-scalars}. At $c_{\max}=2$ we set $J=0$ and solve
the first three equations. The density and velocity equations
describe gravitational growth opposed by pressure gradients; the last
two describe how the pressure changes and is transported. Specifying $\epsilon(a)$
fixes their time-dependent coefficients without specifying the
perturbation $S$.

We start at $a_{\rm in}=10^{-3}$, when $k^2\epsilon\ll1$, and select
the dust growing mode with its associated pressure perturbation.
Writing $\beta=\ud\ln\epsilon/\ud N$, the initial data are
\begingroup
\allowdisplaybreaks[0]
\begin{align}
 \delta_{m,\rm in}&=D_{\rm in}\delta_{L0},
 \qquad \theta_{\rm in}=\delta_{m,\rm in},\notag\\
 S_{\rm in}&=\epsilon_{\rm in}\delta_{m,\rm in}
 \left(1+\frac{2}{2\mu_{\rm in}-1+\beta_{\rm in}}\right),\notag\\
 J_{\rm in}&=\frac{6k^2\epsilon_{\rm in}^2\delta_{m,\rm in}}
 {(2\mu_{\rm in}-1+\beta_{\rm in})
  (3\mu_{\rm in}-2+2\beta_{\rm in})}.
 \label{eq:app-VPT-initial}
\end{align}
\endgroup
The $J$ initial condition is used only at $c_{\max}=3$. These expressions
follow by taking $\delta_m\propto D$, $S\propto\epsilon D$ and
$J\propto\epsilon^2D$ at early times; the factors in their denominators
combine the growth rates with the damping terms in
\eqref{eq:app-cmax3-system}.

The code integrates in $\ln a$, multiplying each right-hand side by
$f=\ud N/\ud\ln a$, using an adaptive eighth-order Runge--Kutta method.
Time derivatives needed for the deceleration multipoles are evaluated
from the same equations. At the end we divide by $\delta_m(k,1)$ to
compare VPT and dust at fixed present-day density, as in
\eqref{eq:main-DV-definition}.

We use the baryon-corrected Eisenstein--Hu no-wiggle spectrum
\cite{Eisenstein:1997ik}, normalised to $\sigma_8=0.8$ in the stated
cosmology, with $T_{\rm CMB}=2.7255\,$K. The variance integrals use
4001 logarithmically spaced wavenumbers from
$k_{\min}=10^{-5}$ to $k_{\max}=1.5\,h\,{\rm Mpc}^{-1}$, including
the cutoff values in table~\ref{tab:radial-cutoffs}. This upper limit
lies below the first zero of $F_{1,\delta}(k,1)$ for every cumulant
order and dispersion normalisation tested, so the division by final
density remains regular. The plotted kernels extend further to show
their high-$k$ behaviour, outside the domain used for variances.
Independent integration in physical moment variables checks the
solutions and the amplitudes in table~\ref{tab:observer-amplitudes}.
The radial integrals use 2401 redshifts over $0\leq z\leq0.2$;
doubling this resolution changes the rms values at $z_{\max}=0.01$
by less than $10^{-8}$ fractionally.

\subsection{Production in overdense regions}
\label{app:production-modulation}

The mean history fixes how much dispersion is produced on average,
but an overdense region may produce more random motion than an
underdense one. To estimate the importance of this omitted effect,
we keep the same $\epsilon(a)$ and add an isotropic production term
$\gamma_\epsilon Q_\epsilon\delta_m$ to the local dispersion equation.
At $c_{\max}=2$ this changes the pressure equation to
\begin{align}
 \frac{\ud S}{\ud N}+2(\mu-1)S
 =3\epsilon\theta+(1+\gamma_\epsilon)Q_\epsilon\delta_m.
 \label{eq:app-production-modulation}
\end{align}
The dimensionless constant $\gamma_\epsilon$ specifies how production
changes with the long-wavelength density: positive values add production
in overdense regions, and negative values reduce it. Our fiducial
calculation has $\gamma_\epsilon=0$. This is a separate sensitivity
test, with no fitted value assigned to $\gamma_\epsilon$.\footnote{We
select the growing initial pressure by replacing the numerator $2$ in
$S_{\rm in}$ in eq.~\eqref{eq:app-VPT-initial} with
$2+\gamma_\epsilon(\beta_{\rm in}+2\mu_{\rm in}-2)$.}
For $\gamma_\epsilon=-1,0,+1$, the acceleration sky rms at
$R=30\,h^{-1}{\rm Mpc}$ is respectively
$3.3\times10^{-7}$, $8.4\times10^{-7}$ and $1.4\times10^{-6}$.
Thus production correlated with the long mode can change its
acceleration by order unity even when the mean dispersion is fixed.
Spatial smoothing does not remove this contribution to the long mode.

\section{Effect of the third cumulant}
\label{app:convergence}

The third cumulant describes an asymmetry of the local velocity
distribution and transports dispersion. We test its importance by
solving eq.~\eqref{eq:app-cmax3-system} with and without $J$, keeping
the mean history, present-day density spectrum and windows fixed.
The mean fourth cumulant is zero in both calculations. This retains
the products of second moments in the fourth central moment,
as shown in eq.~\eqref{eq:fourth-central-moment}.

\begin{figure}[t]
\centering
\includegraphics[width=\textwidth]{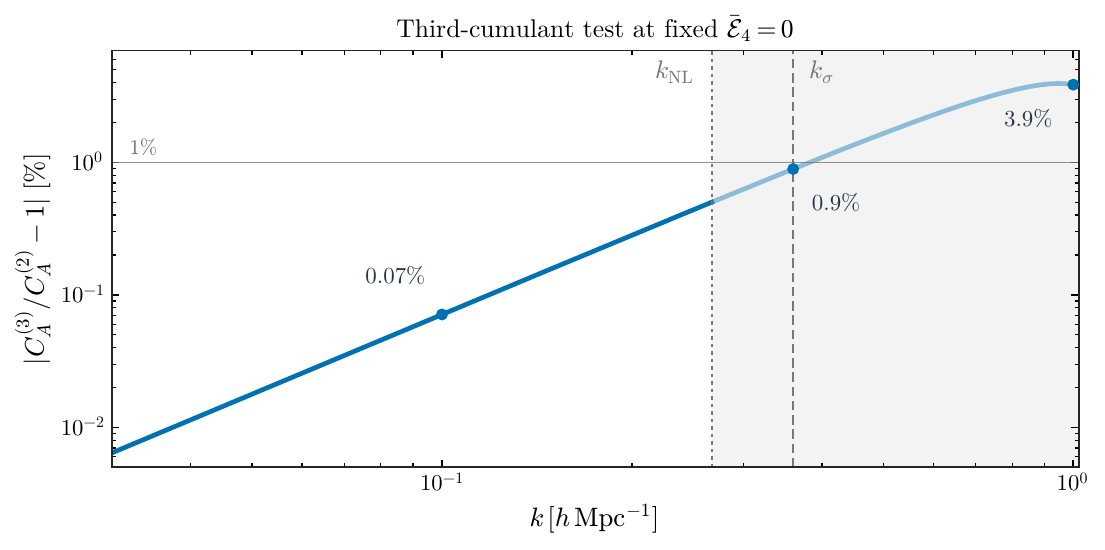}
\caption{Fractional change in $C_A(k,0)$ when the third cumulant is
included. The mean fourth cumulant is set to zero; $\bar{\mathcal E}_4$
in the plot denotes its one-component value divided by the squared
one-component dispersion. The change is
$0.07\%$ at $k=0.1\,h\,{\rm Mpc}^{-1}$, $0.9\%$ at $k_\sigma$ and
$3.9\%$ at $k=1\,h\,{\rm Mpc}^{-1}$. The pale continuation begins
at $k_{\rm NL}$.}
\label{fig:vpt-cumulant-control}
\end{figure}

Figure~\ref{fig:vpt-cumulant-control} shows the fractional change in
$C_A(k,0)$. It reaches about $0.9\%$ at $k_\sigma$ and grows on
shorter scales. The windows suppress most of this difference. For an
rms amplitude $\sigma_w$, we quote the absolute percentage change
\begin{align}
 \Delta_w^{(3,2)}=100\left|
 \frac{\sigma_w^{(3)}}{\sigma_w^{(2)}}-1\right|.
 \label{eq:app-cmax3-shift}
\end{align}
The results are
\begin{center}
\begin{tabular}{lc}
\hline
Window & $\Delta_w^{(3,2)}$ \\
\hline
Gaussian, $R=5\text{--}40\,h^{-1}{\rm Mpc}$ & $\leq0.5\%$ \\
real-space top-hat, $R=5\text{--}40\,h^{-1}{\rm Mpc}$ & $\leq1.3\%$ \\
cumulative, $z_{\max}=0.01$ & $0.4\%$ \\
outer shell, $z_{\max}=0.01$ & $0.8\%$ \\
\hline
\end{tabular}
\end{center}
The local observer multipoles change by less than one percent at both
Gaussian radii (table~\ref{tab:observer-amplitudes}). Their time derivatives
come directly from the evolution equations in appendix~\ref{app:integration}.

This comparison tests the cumulant truncation within linear evolution.
It does not test nonlinear mode coupling, which becomes important at
$k_{\rm NL}=0.267\,h\,{\rm Mpc}^{-1}$, before
$k_\sigma=0.360\,h\,{\rm Mpc}^{-1}$. The Gaussian observer windows
suppress these modes strongly. The top-hat and radial windows retain
more power from them, as quantified in appendix~\ref{app:windows},
and their amplitudes depend on the common upper cutoff.

\section{Acceleration and coarse-grained dynamics}
\label{sec:cg_dust}
\label{app:coarse-grained-dust}

In the effective field theory of large-scale structure, spatial averaging
replaces unresolved motions and gravitational fields by an effective
pressure tensor \cite{Baumann:2010tm,Carrasco:2012cv}. Even a dust flow
with one velocity at each point acquires such a tensor when neighbouring
points with different velocities are averaged together. Its divergence
accelerates the smoothed, mass-weighted flow. Our local Vlasov pressure
instead comes from different particle velocities at the same point.
Spatially averaging that Vlasov flow would add the unresolved bulk-motion
and gravitational contributions as well.

For a sharp comoving cutoff $\Lambda$, let the smoothed velocity be the
smoothed momentum density divided by the smoothed mass density. Its
acceleration relative to the resolved gravitational field is
$A_{\Lambda,i}=-(a\rho_{m,\Lambda})^{-1}
\partial^j\tau^{\rm eff}_{ij}$. The part correlated with a long density
mode has the form
\begin{align}
 A_{\Lambda,i}(\bm k,a)
 =-\frac{{\rm i}k_i}{a}C_{A,\Lambda}(a)\delta_\Lambda(\bm k,a)
 \left[1+O\left(\frac{k^2}{\Lambda^2}\right)\right].
 \label{eq:main-cg-acceleration}
\end{align}
Thus $C_{A,\Lambda}$ enters the long-mode Euler equation as an effective
sound-speed coefficient. In EFT it includes short-scale physics that must
be matched to a nonlinear calculation or data. We compare only the
perturbative contribution from linear short modes. The separation scale
$\Lambda^{-1}$ is independent of the observational smoothing radius $R$.

The required one-loop result is the ultraviolet limit of $P_{13}$ with
Einstein--de Sitter spatial kernels, given in appendix B of
\cite{Nascimento:2024rbv}. Since $P_{13}$ is twice the cross-correlation
of the linear and third-order density fields, its contribution from
$q>\Lambda$ corresponds to
\begin{align}
 \delta_{\rm UV}(\bm k,a)
 =-\frac{61}{630}k^2D^3\delta_{L0}(\bm k)
 \int_\Lambda^\infty\frac{\ud q}{2\pi^2}P_{L0}(q).
 \label{eq:cg-density-matching}
\end{align}
Here $P_{L0}$ is the pressureless linear spectrum extrapolated to today.
To recover the acceleration producing this density correction, we apply
the $\Lambda$CDM density evolution operator, with $x=\ln a$:
\begin{align}
 \mathcal L_x
 &=\frac{\ud^2}{\ud x^2}
 +\left(2-\frac32\Omega_m\right)\frac{\ud}{\ud x}
 -\frac32\Omega_m,
 &\mathcal L_xD^3&=3(2f^2+\Omega_m)D^3.
 \label{eq:cg-growth-operator}
\end{align}
The second identity follows by differentiating $D^3$ and using
$\mathcal L_xD=0$. Equating
$\mathcal L_x\delta_{\rm UV}=-k^2 C_{A,\Lambda}^{(2)}D\delta_{L0}/
\mathcal H^2$ gives
\begin{align}
 C_{A,\Lambda}^{(2)}
 &=\frac{61}{210}\left(2+\frac{\Omega_m}{f^2}\right)I_v,
 &I_v&=(\mathcal Hf)^2D^2
 \int_\Lambda^\infty\frac{\ud q}{2\pi^2}P_{L0}(q).
 \label{eq:main-cg-coefficient}
\end{align}
The superscript denotes the contribution quadratic in short-mode
amplitude. Here $I_v$ is the three-dimensional variance of their linear
bulk velocity. This expression keeps $\Lambda$CDM growth with
Einstein--de Sitter spatial kernels and reduces to $61I_v/70$ in
Einstein--de Sitter.

At $\Lambda=k_{\rm NL}(0)\simeq0.27\,h\,{\rm Mpc}^{-1}$ we obtain
\begin{align}
 C_{A,\Lambda}^{(2)}(0)
 &\simeq2.36\times10^4\,({\rm km}\,{\rm s}^{-1})^2,
 &C_A(k\to0,0)
 &\simeq2.24\times10^4\,({\rm km}\,{\rm s}^{-1})^2.
 \label{eq:main-cg-comparison}
\end{align}
The perturbative coarse-grained value is $5.4\%$ larger than the
fiducial VPT value. Raising $\Lambda$ to $k_\sigma(0)$ lowers it to
$1.7\times10^4\,({\rm km}\,{\rm s}^{-1})^2$. This agreement concerns
the specified linear short-mode estimate. The full nonlinear coefficient
requires the kinetic and gravitational contributions of the unresolved
structures, including their virial cancellation \cite{Baumann:2010tm}.

\section{Observer multipoles and amplitudes}
\label{app:observer-normalisation}
\subsection{Multipole transfers}
\label{app:observer-transfers}

Pressure changes the observer multipoles in two linked ways. It alters
the growth of density and velocity, changing the expansion and shear
terms already present for dust, and it adds terms containing the
acceleration explicitly. We label these contributions \emph{grow} and
$A$, respectively, as in section~\ref{sec:deceleration-multipoles} and
figure~\ref{fig:vpt-Q-transfers}. We keep them separate to identify where
the change enters the cosmographic expression, then add them to obtain
the total correction.

Writing the VPT--dust deceleration multipoles as
\begin{align}
 \Delta\mathbb Q_i
 &= {\rm i}\hat k_i
 \big(T_{1,{\rm grow}}+T_{1,A}\big)\delta_0,
 \notag\\
 \Delta\mathbb Q_{ij}
 &=\hat k_{\langle i}\hat k_{j\rangle}
 \big(T_{2,{\rm grow}}+T_{2,A}\big)\delta_0,
 \notag\\
 \Delta\mathbb Q_{ijk}
 &= {\rm i}\hat k_{\langle i}\hat k_j\hat k_{k\rangle}
 T_{3,{\rm grow}}\delta_0,
 \label{eq:observer-Q123-transfers}
\end{align}
gives, at $a=1$,
\begin{align}
 T_{1,{\rm grow}}
 &=-\frac{3ck}{5\mathcal H_0}\Delta G,
 &
 T_{1,A}
 &=-\left(\frac{\partial}{\partial\ln a}+1+q\right)T_A,
 \notag\\
 T_{2,{\rm grow}}
 &=\frac{\partial\Delta G}{\partial\ln a}+(1+q)\Delta G,
 &
 T_{2,A}
 &=-\frac{ck}{\mathcal H_0}T_A,
 \notag\\
 T_{3,{\rm grow}}
 &=-\frac{ck}{\mathcal H_0}\Delta G.
 \label{eq:observer-Q123-components}
\end{align}

Here $\Delta G=\partial\mathcal D/\partial\ln a-fD$ is the difference
between VPT and dust velocity growth at the same final density.
The octupole contains only a shear gradient, so its correction is entirely
in the \emph{grow} contribution. The scalar transfers
satisfy
\begin{align}
 T_{3,{\rm grow}}=\frac53T_{1,{\rm grow}},
 \label{eq:observer-Q13-relation}
\end{align}
as follows from their common velocity-growth source.

For the volume-weighted cumulative sample, we replace $z$ in the local
Taylor expression by $\bar z_w=\int\ud z\,w(z)z$. Define
${\cal T}_{\eta,\ell}$ as the scalar coefficient transferring $\delta_0$
to $(\ln10)\Delta\eta_{{\cal A}_\ell}$, with the PSTF tensor and the
factor ${\rm i}$ for odd $\ell$ displayed as in
\eqref{eq:observer-Q123-transfers}. The coefficients are
\begin{align}
 {\cal T}_{\eta,1}
 &=T_A+\frac{\bar z_w}{2}
   \big(T_{1,{\rm grow}}+T_{1,A}\big),
 \notag\\
 {\cal T}_{\eta,2}
 &=-\Delta G+\frac{\bar z_w}{2}
   \big(T_{2,{\rm grow}}+T_{2,A}\big),
 \notag\\
 {\cal T}_{\eta,3}
 &=\frac{\bar z_w}{2}T_{3,{\rm grow}}.
 \label{eq:observer-eta123-transfers}
\end{align}

The signed contributions are combined before taking their rms, retaining
their cross-correlations. The dust transfers use the pressureless
growth factor $D(a)$, with $D(1)=1$; the shear transfer is $-fD$.
The deceleration dipole and octupole have scalar transfers
$-3ckfD/(5\mathcal H)$ and $-ckfD/\mathcal H$. The dust quadrupole is
\begin{align}
 T_{2,{\rm dust}}=D\left(\frac32\Omega_m+2qf\right),
 \label{eq:observer-dust-Q2}
\end{align}
using the pressureless growth equation. Its two terms partly cancel
today, giving about $-0.20$. This small dust coefficient explains why
the fractional change in the deceleration quadrupole can be appreciably
larger than the fractional change in the Hubble shear.

\subsection{Angular normalisation}

For the scalar PSTF basis, contracting one index with $\hat k^a$
lowers the rank and multiplies the result by $\ell/(2\ell-1)$
(appendix~\ref{app:velocity-cumulants}; \cite{GebbieEllis:2000}).
Trace terms vanish against a PSTF tensor, so repeated contraction gives
\begin{align}
 \hat k_{\langle{\cal A}_\ell\rangle}
 \hat k^{\langle{\cal A}_\ell\rangle}
 =\prod_{r=1}^{\ell}\frac{r}{2r-1}
 =\frac{\ell!}{(2\ell-1)!!}.
 \label{eq:pstf-unit-norm}
\end{align}
The norms are $1$, $2/3$ and $2/5$ for ranks one to three. Combining
this contraction with the sky average
\eqref{eq:observer-sky-rms} gives
\begin{align}
\sigma_{\Delta\eta,\ell}^2
 =\frac{(\ell!)^2}{(2\ell+1)!!(2\ell-1)!!\,\ln^2 10}
 \int\ud\ln k\,\Delta_0^2(k)
 \left|{\cal T}_{\eta,\ell}(k)\right|^2 W_{\rm G}^2(kR).
 \label{eq:observer-eta-multipole-rms}
\end{align}

The corresponding mean-square tensor norm is
\begin{align}
 \left\langle\Delta\eta_{{\cal A}_\ell}\Delta\eta^{{\cal A}_\ell}\right\rangle
 =\frac{\ell!}{(2\ell-1)!!\,\ln^2 10}
 \int\ud\ln k\,\Delta_0^2(k)
 \left|{\cal T}_{\eta,\ell}(k)\right|^2
 W_{\rm G}^2(kR).
 \label{eq:observer-tensor-variance}
\end{align}

Thus the tensor rms is converted to sky rms by
$\sqrt{\ell!/(2\ell+1)!!}$. Together with
eq.~\eqref{eq:observer-Q13-relation}, this gives the growth-octupole to
growth-dipole sky-rms ratio $\sqrt{4/21}$.
For the leading acceleration, the conversion is
\begin{align}
\sigma_{\eta_A,1}=\frac{\sigma_{d,R}}{\sqrt3\,\ln10},
\label{eq:window-amplitude-conversion}
\end{align}
where $\sigma_{d,R}$ is the rms magnitude of the smoothed acceleration
vector and the spatial window must be the same on both sides.

\subsection{Numerical amplitudes and resolution}
\label{app:numerical-amplitudes}

Table~\ref{tab:observer-amplitudes} uses Gaussian radii of
$30\,h^{-1}{\rm Mpc}$ and $13.4\,h^{-1}{\rm Mpc}$. The finer radius
comes from matching a top-hat extending to $z=0.01$. Expanding the two
windows at small $k$ gives
\begin{align}
 W_{\rm G}=1-\frac12k^2R_{\rm G}^2+O(k^4),\qquad
 W_{\rm TH}=1-\frac1{10}k^2R_{\rm TH}^2+O(k^4),\qquad
 R_{\rm G}=\frac{R_{\rm TH}}{\sqrt5}.
 \label{eq:window-second-moment-match}
\end{align}
With $R_{\rm TH}=\chi(0.01)=29.9\,h^{-1}{\rm Mpc}$, this gives
$R_{\rm G}=13.38\,h^{-1}{\rm Mpc}$, quoted as $13.4$ throughout.
The matching fixes their large-scale smoothing; their short-wavelength
tails still differ. The local $O(z)$ terms use the cumulative volume
weight, for which $\bar z_w=7.50\times10^{-3}$.

\begin{table}[t]
\centering
\small
\setlength{\tabcolsep}{4pt}
\begin{tabular}{lrrrrr}
\hline
Contribution & \shortstack{Tensor rms\\in $\Delta\eta$} & \shortstack{Sky rms\\in $\Delta\eta$}
 & \shortstack{Dust\\sky rms} & \shortstack{Fraction\\$[\%]$}
 & \shortstack{$c_{\max}:2\to3$\\$[\%]$}\\[2pt]
\hline
\multicolumn{6}{c}{$R_{\rm G}=30\,h^{-1}{\rm Mpc}$}\\
\hline
Hubble shear, $z^0$ & $6.16\times10^{-5}$ & $2.25\times10^{-5}$ & $7.50\times10^{-3}$ & 0.30 & 0.015\\
Acceleration, $z^0$ & $1.46\times10^{-6}$ & $8.42\times10^{-7}$ & $0$ & --- & 0.016\\
Acceleration correction, $z$ & $2.14\times10^{-9}$ & $1.24\times10^{-9}$ & $0$ & --- & 0.077\\
Growth dipole, $z$ & $2.92\times10^{-5}$ & $1.69\times10^{-5}$ & $3.52\times10^{-3}$ & 0.48 & 0.020\\
Deceleration quadrupole, $z$ & $9.85\times10^{-7}$ & $3.60\times10^{-7}$ & $1.24\times10^{-5}$ & 2.91 & 0.026\\
Octupole, $z$ & $3.08\times10^{-5}$ & $7.36\times10^{-6}$ & $1.54\times10^{-3}$ & 0.48 & 0.020\\
Full dipole, through $z$ & $3.04\times10^{-5}$ & $1.76\times10^{-5}$ & $3.52\times10^{-3}$ & 0.50 & 0.020\\
Full quadrupole, through $z$ & $6.06\times10^{-5}$ & $2.21\times10^{-5}$ & $7.51\times10^{-3}$ & 0.29 & 0.015\\
\hline
\multicolumn{6}{c}{$R_{\rm G}=13.4\,h^{-1}{\rm Mpc}$}\\
\hline
Hubble shear, $z^0$ & $6.14\times10^{-4}$ & $2.24\times10^{-4}$ & $1.75\times10^{-2}$ & 1.28 & 0.069\\
Acceleration, $z^0$ & $6.87\times10^{-6}$ & $3.97\times10^{-6}$ & $0$ & --- & 0.072\\
Acceleration correction, $z$ & $9.83\times10^{-9}$ & $5.67\times10^{-9}$ & $0$ & --- & 0.339\\
Growth dipole, $z$ & $6.26\times10^{-4}$ & $3.61\times10^{-4}$ & $1.66\times10^{-2}$ & 2.17 & 0.093\\
Deceleration quadrupole, $z$ & $9.77\times10^{-6}$ & $3.57\times10^{-6}$ & $2.89\times10^{-5}$ & 12.36 & 0.119\\
Octupole, $z$ & $6.60\times10^{-4}$ & $1.58\times10^{-4}$ & $7.25\times10^{-3}$ & 2.17 & 0.093\\
Full dipole, through $z$ & $6.31\times10^{-4}$ & $3.65\times10^{-4}$ & $1.66\times10^{-2}$ & 2.19 & 0.093\\
Full quadrupole, through $z$ & $6.04\times10^{-4}$ & $2.21\times10^{-4}$ & $1.76\times10^{-2}$ & 1.26 & 0.068\\
\hline

\end{tabular}
\caption{Local VPT--dust corrections at fixed present-day density, using
$c_{\max}=3$. Tensor and sky rms are defined in
\eqref{eq:observer-tensor-variance} and
\eqref{eq:observer-eta-multipole-rms}. ``Fraction'' is the correction's
sky rms as a percentage of the dust sky rms; the last column gives the
change on including the third cumulant. Terms through $O(z)$ use
$\bar z_w=0.0075$. Full multipoles add the signed contributions before
taking the rms.}
\label{tab:observer-amplitudes}
\end{table}

For each contribution with transfer $\mathcal T(k)$, the
rms value of the spatial expansion parameter $k\chi$ is
\begin{align}
 \big[k\chi(\bar z_w)\big]_{\rm rms}^2
 =\chi^2(\bar z_w)
 \frac{\int\ud\ln k\,k^2\Delta_0^2|\mathcal T|^2W_{\rm G}^2}
 {\int\ud\ln k\,\Delta_0^2|\mathcal T|^2W_{\rm G}^2}.
 \label{eq:observer-kchi}
\end{align}
The same variance weights determine the fraction above $k_{\rm NL}$.
The variance-weighted rms of $k\chi(\bar z_w)$ ranges from
$1.1$ to $1.5$ at $30\,h^{-1}{\rm Mpc}$ and from $2.2$ to $3.2$
at $R=13.4\,h^{-1}{\rm Mpc}$. The corresponding dust transfers give
$0.81$--$1.07$ and $1.63$--$2.25$ at the same two radii.
Modes above $k_{\rm NL}$ supply less than $0.1\%$ of the observer
variances, so the gradient limitation is distinct from direct
short-wavelength contamination of these local coefficients.

The growth correction to the deceleration dipole follows from its
$O(z)$ contribution to $\eta$ by multiplying by $2\ln10/\bar z_w$.
At the two radii, its sky rms is $0.010$ and $0.22$, and its rms vector
magnitude is $0.018$ and $0.38$. In an isotropic ensemble the rms
component along a specified axis equals the sky rms. This fixed-axis
quantity is the comparison used in section~\ref{sec:intrinsic-dipole}.
At $R=13.4\,h^{-1}{\rm Mpc}$ the full dipole correction has sky rms
$3.6\times10^{-4}$, rising to $7.6\times10^{-4}$ at the largest tested
dispersion normalisation: about $3\%$ and $6\%$ of the measured CF4a
sky rms. These use a cumulative weight ending where the CF4a sample
begins, so they compare amplitude scales rather than identical estimators.
At $R_{\rm G}=30\,h^{-1}{\rm Mpc}$ the opposing deceleration term
reduces the shear quadrupole rms by $1.6\%$, and the explicit
acceleration correction is $1.5\times10^{-3}$ of the leading dipole.

\subsection{Cosmic variance}
\label{app:observer-probability}

We use the chi distributions described in
section~\ref{sec:observer-variation}, with the angular normalisation
of eq.~\eqref{eq:observer-sky-rms}. Figure~\ref{fig:vpt-observer-multipoles}
shows probability density per logarithmic interval in sky amplitude,
with shading for the central $68\%$ interval.

The two dipoles have transfers
${\cal T}_{\eta,1,A}=T_A+\bar z_wT_{1,A}/2$ and
${\cal T}_{\eta,1,{\rm grow}}=\bar z_wT_{1,{\rm grow}}/2$,
as in eq.~\eqref{eq:observer-eta123-transfers}. Their covariance is obtained
by replacing $|{\cal T}_{\eta,1}|^2$ in
\eqref{eq:observer-eta-multipole-rms} with the product of these two
transfers. At $R=30\,h^{-1}{\rm Mpc}$ the correlation coefficient is
$0.83$, giving
\begin{align}
 {\rm Prob}\!\left(
  |\boldsymbol\eta_A|>|\boldsymbol\eta_{1,{\rm grow}}|
 \right)=3.72\times10^{-5}.
 \label{eq:observer-dominance}
\end{align}

Observers with $|\boldsymbol\eta_A|>|\boldsymbol\eta_{1,{\rm grow}}|$ mainly have
a small growth dipole. The $99$th percentile of
the acceleration sky amplitude is $1.6\times10^{-6}$, compared with
the growth-dipole sky rms $1.7\times10^{-5}$. 

\section{Spatial and radial windows}
\label{app:windows}

A window determines which spatial scales contribute to a quoted
acceleration amplitude. We first smooth the field around the observer
and compare it with the bulk velocity on the same scale. We then
average the radial acceleration evaluated at source positions, where
changes in direction and sign along the line of sight can cancel.
These two calculations explain the spatial and depth dependence in
figures~\ref{fig:vpt-smoothing-windows}--\ref{fig:vpt-dipole-windows}.
The second is a test of the acceleration field's spatial variation;
a distance prediction also needs the remaining light-cone terms.

\subsection{Observer smoothing}
\label{app:vpt-fixed-smoothing}

For an isotropic spatial window $W(kR)$, the smoothed acceleration obeys
\begin{align}
 \big\langle d_i(R,z)d_j(R,z)\big\rangle
 &=\frac13\delta_{ij}\sigma_d^2(R,z),
 \notag\\
 \sigma_d^2(R,z)&=\int_{k_{\min}}^{k_{\max}}\ud\ln k\,
 \Delta_0^2(k)T_A^2(k,z)W^2(kR).
 \label{eq:app-sigma-d-R}
\end{align}

Thus $\sigma_d$ is the rms magnitude of the three-dimensional vector,
and each Cartesian component has variance $\sigma_d^2/3$. We use
\begin{align}
 W_{\rm G}(x)=e^{-x^2/2},
 \qquad
 W_{\rm TH}(x)=\frac{3j_1(x)}{x}.
 \label{eq:app-windows}
\end{align}

The second-moment match in eq.~\eqref{eq:window-second-moment-match}
associates the Gaussian radius $30\,h^{-1}{\rm Mpc}$ with a top-hat
radius $67\,h^{-1}{\rm Mpc}$. Their Fourier tails differ.

\begin{figure}[t]
\centering
\includegraphics[width=\textwidth]{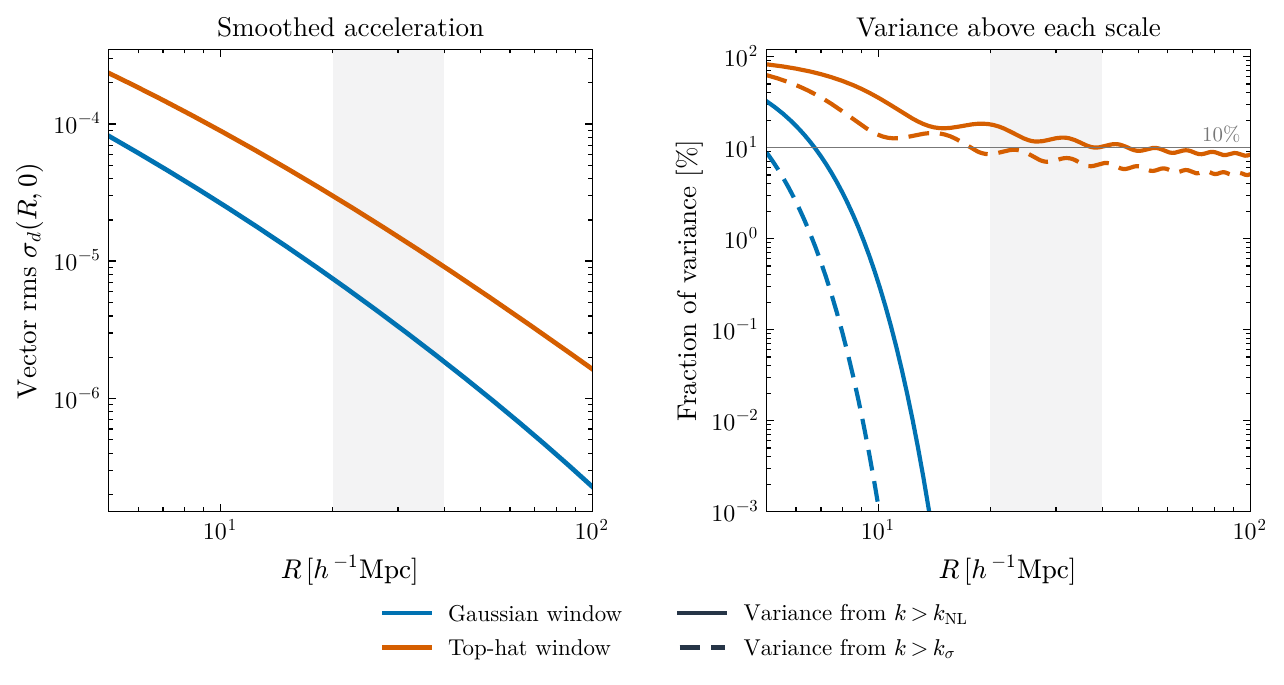}
\caption{Present-day acceleration after spatial smoothing.
Left: three-dimensional vector rms
$\sigma_d=\langle|\bm A/(cH)|^2\rangle^{1/2}$ for Gaussian and real-space
top-hat windows. Divide by $\sqrt3\ln10$ for sky rms in $\eta$. The shaded region marks $R=20$--$40\,h^{-1}{\rm Mpc}$;
equal radii in the two windows represent different resolutions.
Right: fractions of the $c_{\max}=2$ variance above $k_{\rm NL}$
(solid) and $k_\sigma$ (dashed). The horizontal line marks $10\%$.}
\label{fig:vpt-smoothing-windows}
\end{figure}

Figure~\ref{fig:vpt-smoothing-windows} shows the acceleration amplitude
and the variance fractions above $k_{\rm NL}$ and $k_\sigma$.

For Gaussian radii $R=10$--$40\,h^{-1}{\rm Mpc}$, less than $0.4\%$
of the acceleration variance comes directly from $k>k_{\rm NL}$.
The top-hat retains more short-wavelength power: the corresponding
fraction is $10$--$18\%$ at $R=20$--$40\,h^{-1}{\rm Mpc}$,
$35\%$ at $R=10\,h^{-1}{\rm Mpc}$ and $82\%$ at
$R=5\,h^{-1}{\rm Mpc}$. The small-radius top-hat estimates depend
strongly on the linear treatment of those modes.

\begin{table}[t]
\centering
\begin{tabular}{ccc}
\hline
$R\,[h^{-1}{\rm Mpc}]$ & Gaussian vector rms $\sigma_d$ & Top-hat vector rms $\sigma_d$\\
\hline
5 & $8.19\times10^{-5}$ & $2.35\times10^{-4}$\\
10 & $2.64\times10^{-5}$ & $8.91\times10^{-5}$\\
20 & $7.49\times10^{-6}$ & $3.00\times10^{-5}$\\
40 & $1.85\times10^{-6}$ & $9.09\times10^{-6}$\\
\hline
\end{tabular}
\caption{Spatially smoothed acceleration-vector rms at $z=0$ and
$c_{\max}=2$. Dividing by $\sqrt3\ln10$ gives sky rms in $\eta$.}
\label{tab:spatial-amplitudes}
\end{table}

The bulk velocity on the same spatial scale follows from the subhorizon
continuity equation, which gives $\vartheta=-\delta_m'$ and hence
\begin{align}
v_i(\bm k,1)
={\rm i}\hat k_i\frac{\mathcal H_0}{k}
\left.\frac{\partial\mathcal D(k,a)}{\partial\ln a}\right|_{a=1}
\delta_0(\bm k).
\label{eq:main-velocity-transfer}
\end{align}

For top-hat smoothing, the two variances are
\begin{align}
\sigma_{d,R}^2
&=\int_{k_{\min}}^{k_{\max}}\ud\ln k\,
\Delta_0^2(k)T_A^2(k,0)W_{\rm TH}^2(kR),
\\
\left(\frac{V_R}{c}\right)^2
&=\int_{k_{\min}}^{k_{\max}}\ud\ln k\,
\Delta_0^2(k)
\left[\frac{\mathcal H_0}{ck}
\left.\frac{\partial\mathcal D(k,a)}{\partial\ln a}\right|_{a=1}
\right]^2
W_{\rm TH}^2(kR).
\label{eq:main-field-smoothing}
\end{align}

\begin{figure}[t]
\centering
\includegraphics[width=0.78\textwidth]{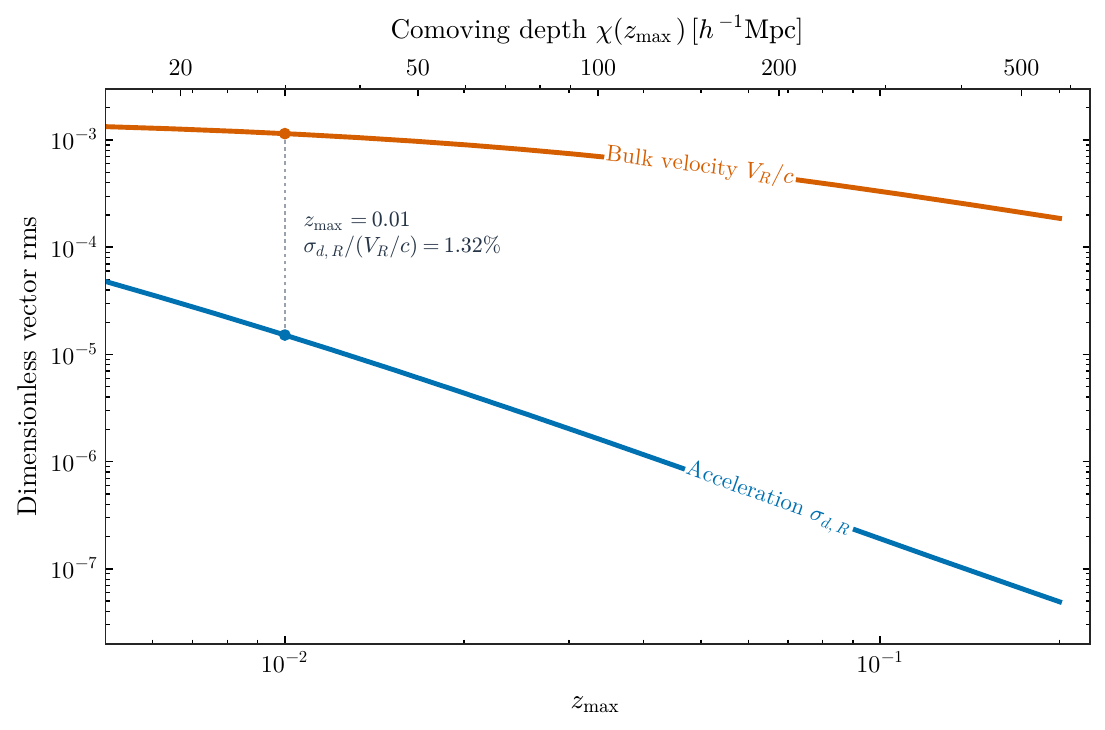}
\caption{Present-day matter fields after real-space top-hat smoothing on
$R=\chi(z_{\max})$. The orange curve is the rms bulk velocity $V_R/c$ and
the blue curve is the rms dimensionless acceleration $\sigma_{d,R}$. These
are three-dimensional vector rms values; the acceleration contribution
to sky rms in $\eta$ is $\sigma_{d,R}/(\sqrt3\ln10)$. Their ratio is approximately $1.3\%$ at
$z_{\max}\simeq0.01$, marked by the filled circles.}
\label{fig:vpt-coherent-fields}
\end{figure}

In figure~\ref{fig:vpt-coherent-fields} we set $R=\chi(z_{\max})$ to
compare these two fields on a scale set by survey depth.
At $z_{\max}=0.01$,
\begin{align}
\frac{V_R}{c}&=1.15\times10^{-3},
&
\sigma_{d,R}&=1.52\times10^{-5},
&
\frac{\sigma_{d,R}}{V_R/c}&=1.32\%.
\label{eq:app-reference-field-values}
\end{align}

The acceleration-vector rms corresponds to sky rms
$3.8\times10^{-6}$ in $\eta$ by
eq.~\eqref{eq:window-amplitude-conversion}. The Gaussian value at
$R_{\rm G}=30\,h^{-1}{\rm Mpc}$ is smaller because it filters
more broadly than a top-hat with the same numerical radius.
The Gaussian with $R=13.4\,h^{-1}{\rm Mpc}$ gives $4.0\times10^{-6}$, close to the
top-hat value despite their different Fourier tails.

The corresponding Doppler dipole is enhanced by the inverse source
redshift. At the outer radius, its characteristic amplitude is
\begin{align}
 \Xi_R\equiv\frac{V_R}{H_0R}\simeq\frac{V_R/c}{z_{\max}},
 \qquad R=\chi(z_{\max}).
 \label{eq:main-hubble-flow-ratio}
\end{align}
At $z_{\max}=0.01$, $\Xi_R=0.115$, so the acceleration-to-Doppler
ratio is only $z_{\max}\sigma_{d,R}/(V_R/c)\simeq1.3\times10^{-4}$.
Here the smoothing radius changes with depth. At fixed radius the
intrinsic contribution instead falls relative to Doppler in direct
proportion to $z$, as discussed in section~\ref{sec:covariant-start}.

\subsection{Source-position projection}
\label{app:source-position-projection}
\label{sec:sample-depth}

\begin{figure}[t]
\centering
\includegraphics[width=0.78\textwidth]{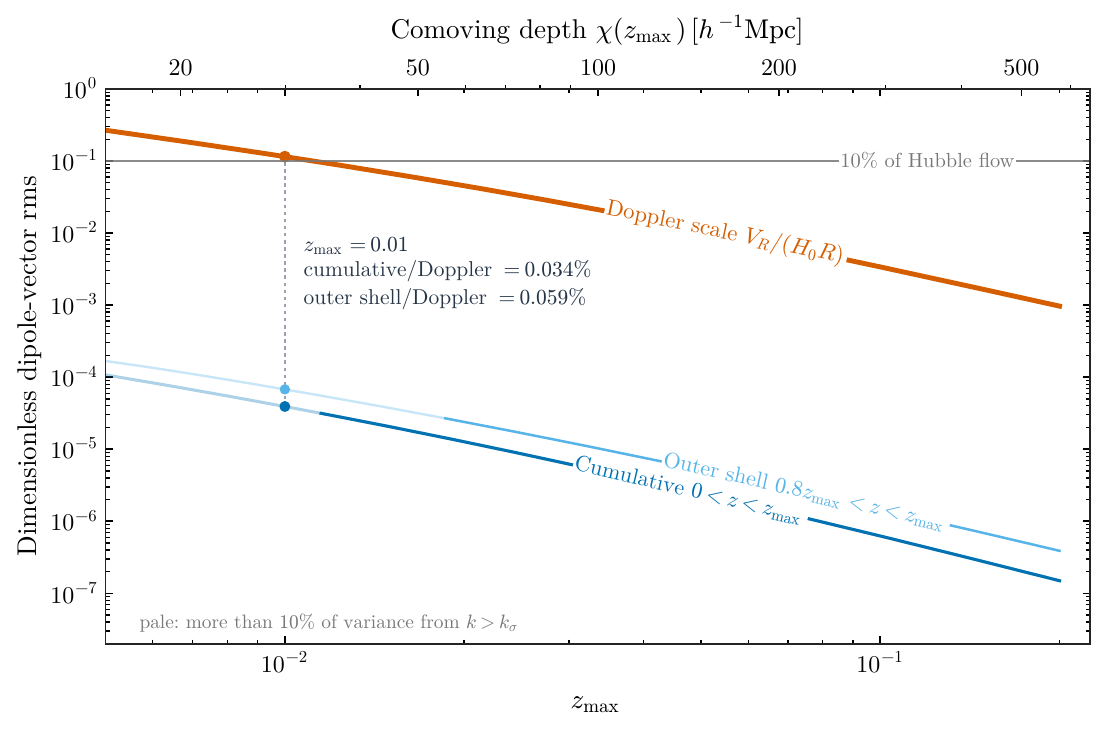}
\caption{Acceleration dipole-vector rms $\sigma_{d,w}$ for sources
throughout the volume and in the outer shell
$0.8z_{\max}<z<z_{\max}$. The orange curve is the bulk-flow Doppler
scale $\Xi_R=V_R/(H_0R)$, with $R=\chi(z_{\max})$.
Filled circles mark $z_{\max}=0.01$; the horizontal line marks
$\Xi_R=0.1$. Pale acceleration curves receive more than $10\%$ of
their variance from $k>k_\sigma$. Divide $\sigma_{d,w}$ by
$\sqrt3\ln10$ to obtain sky rms in $d_\parallel/\ln10$.}
\label{fig:vpt-dipole-windows}
\end{figure}

Figure~\ref{fig:vpt-dipole-windows} compares the acceleration dipole
averaged over all sources within a depth with that from sources in an
outer shell. The broader average allows more cancellation between
nearby and distant regions. We evaluate
$d_\parallel(z,\bm n)=\bm n\cdot\bm d[\chi(z)\bm n,z]$ at the source
positions, using a normalised redshift weight $w(z)$.
The dipole vector of this sky pattern is
\begin{align}
 \bm d_w=\frac{3}{4\pi}\int\ud\Omega_{\bm n}\,\bm n
 \int\ud z\,w(z)d_\parallel(z,\bm n),
 \qquad \int\ud z\,w(z)=1.
 \label{eq:app-radial-dipole-vector}
\end{align}
The factor $3$ recovers $\bm d$ from a constant vector field.
For a longitudinal Fourier mode, the angular integral gives
$3{\rm i}\hat{\bm k}j_1'[k\chi]$: the radial projection supplies the
Bessel derivative. Inserting the transfer $T_A$ and averaging over the
density field therefore gives
\begin{align}
 \sigma_{d,w}^2\equiv\langle|\bm d_w|^2\rangle
 =9\int_{k_{\min}}^{k_{\max}}\ud\ln k\,\Delta_0^2(k)
 \left|\int\ud z\,w(z)T_A(k,z)j_1'[k\chi(z)]\right|^2.
 \label{eq:main-source-radial-variance}
\end{align}
The radial integral is taken before squaring, so contributions from
different distances can cancel. Since $j_1'(0)=1/3$, a window at the
observer recovers the unsmoothed vector variance. The sky rms of the
projected dipole in $d_\parallel/\ln10$ is
$\sigma_{d,w}/(\sqrt3\ln10)$.

For constant comoving source density and equal source weights, the
number of sources in a redshift interval is proportional to
$\chi^2/H$. We therefore use
\begin{align}
 w_V(z;z_1,z_2)=
 \frac{\chi^2(z)/H(z)}
 {\displaystyle\int_{z_1}^{z_2}\ud z'\,\chi^2(z')/H(z')},
 \qquad z_1<z<z_2,
 \label{eq:app-volume-window}
\end{align}
with zero weight outside this interval. The cumulative sample uses
$(z_1,z_2)=(0,z_{\max})$ and the outer shell uses
$(0.8z_{\max},z_{\max})$.

At $z_{\max}=0.01$, their dipole-vector rms values are
$3.92\times10^{-5}$ and $6.78\times10^{-5}$, respectively.
The outer shell has less radial cancellation and receives $59\%$ of
its variance from $k>k_{\rm NL}$, compared with $16\%$ for the
cumulative sample. These estimates consequently depend on the
extrapolation of linear evolution to short scales. Both projected
acceleration amplitudes remain far below the bulk-flow Doppler scale
$\Xi_R\simeq0.1$ shown in the figure.

\subsection{Ultraviolet cutoff}

The radial calculation uses $k_{\max}=1.5\,h\,{\rm Mpc}^{-1}$.
Table~\ref{tab:radial-cutoffs} shows its dependence on this cutoff at
$z_{\max}=0.01$. The cumulative projection changes modestly above
$k_\sigma$, whereas the outer-shell value changes substantially because
it retains more short-wavelength power.
The top-hat cumulant comparison is regulated at the same cutoff in
appendix~\ref{app:convergence}.

\begin{table}[t]
\centering
\begin{tabular}{crr}
\hline
$k_{\max}\ [h\,{\rm Mpc}^{-1}]$ & \shortstack{Cumulative\\vector rms $\sigma_{d,w}$} & \shortstack{Outer-shell\\vector rms $\sigma_{d,w}$}\\
\hline
0.267 & $3.59\times10^{-5}$ & $4.35\times10^{-5}$\\
0.36 & $3.66\times10^{-5}$ & $5.25\times10^{-5}$\\
0.5 & $3.81\times10^{-5}$ & $6.10\times10^{-5}$\\
0.75 & $3.89\times10^{-5}$ & $6.69\times10^{-5}$\\
1 & $3.91\times10^{-5}$ & $6.76\times10^{-5}$\\
1.5 & $3.92\times10^{-5}$ & $6.78\times10^{-5}$\\

\hline
\end{tabular}
\caption{Cutoff dependence of source-position dipole-vector rms for
$z_{\max}=0.01$, using $c_{\max}=2$. The lowest cutoff is $k_{\rm NL}$
and the second is $k_\sigma$. Converting either column to sky rms of
$d_\parallel/\ln10$ divides it by $\sqrt3\ln10$.}
\label{tab:radial-cutoffs}
\end{table}

\providecommand{\href}[2]{#2}\begingroup\raggedright\endgroup

\end{document}